\documentclass[%
 reprint,
 superscriptaddress,
 amsmath,amssymb,
 aps,
]{revtex4-2}
\usepackage{graphicx}
\usepackage{dcolumn}
\usepackage{bm}
\usepackage{tabularx,colortbl} 
\usepackage{multirow}
\usepackage{array}
\usepackage{afterpage}
\usepackage{color}
\definecolor{lightGray}{rgb}{0.863, 0.863, 0.863}
\newcolumntype{Y}{>{\centering\arraybackslash}X} 

\usepackage{hyperref}

\usepackage{setspace}

\def\Isat{I_\mathrm{sat}}
\providecommand{\abs}[1]{\lvert#1\rvert}
\usepackage[titletoc]{appendix}

\begin{document}


\title{Photon emission correlation spectroscopy as an analytical tool for quantum defects}

\author{Rebecca E. K. Fishman}
\affiliation{ 
Quantum Engineering Laboratory, Department of Electrical and Systems Engineering, University of Pennsylvania, 200 S. 33rd St. Philadelphia, Pennsylvania, 19104, USA
}
\affiliation{
Department of Physics and Astronomy, University of Pennsylvania, 209 S. 33rd St. Philadelphia, Pennsylvania 19104, USA
}

\author{Raj N. Patel}
\affiliation{ 
Quantum Engineering Laboratory, Department of Electrical and Systems Engineering, University of Pennsylvania, 200 S. 33rd St. Philadelphia, Pennsylvania, 19104, USA
}

\author{David A. Hopper}
\thanks{Present address:
MITRE Corporation, 7515 Colshire Dr.
McLean, VA 22102, USA
}

\affiliation{ 
Quantum Engineering Laboratory, Department of Electrical and Systems Engineering, University of Pennsylvania, 200 S. 33rd St. Philadelphia, Pennsylvania, 19104, USA
}
\affiliation{
Department of Physics and Astronomy, University of Pennsylvania, 209 S. 33rd St. Philadelphia, Pennsylvania 19104, USA
}

\author{Tzu-Yung Huang}
\affiliation{ 
Quantum Engineering Laboratory, Department of Electrical and Systems Engineering, University of Pennsylvania, 200 S. 33rd St. Philadelphia, Pennsylvania, 19104, USA
}

\author{Lee C. Bassett}
\email[Corresponding author.  Email: ]{lbassett@seas.upenn.edu}
\affiliation{ 
Quantum Engineering Laboratory, Department of Electrical and Systems Engineering, University of Pennsylvania, 200 S. 33rd St. Philadelphia, Pennsylvania, 19104, USA
}

\date{\today}

\begin{abstract}
Photon emission correlation spectroscopy is an indispensable tool for the study of atoms, molecules, and, more recently, solid-state quantum defects. 
In solid-state systems, its most common use is as an indicator of single-photon emission, a key property for quantum technology. 
Beyond an emitter's single-photon purity, however, photon correlation measurements can provide a wealth of information that can reveal details about its electronic structure and optical dynamics that are hidden by other spectroscopy techniques.
This tutorial presents a standardized framework for using photon emission correlation spectroscopy to study quantum emitters, including discussion of theoretical background, considerations for data acquisition and statistical analysis, and interpretation.  
We highlight important nuances and best practices regarding the commonly-used $g^{(2)}(\tau=0)<0.5$ test for single-photon emission. 
Finally, we illustrate how this experimental technique can be paired with optical dynamics simulations to formulate an electronic model for unknown quantum emitters, enabling the design of quantum control protocols and assessment of their suitability for quantum information science applications.

\end{abstract}

\maketitle

\setcounter{tocdepth}{2}
\begin{singlespace}
\begin{spacing}{1}
\tableofcontents
\end{spacing}
\end{singlespace}

\begin{figure*}
\includegraphics[trim={0 0 0 0},width = \textwidth]{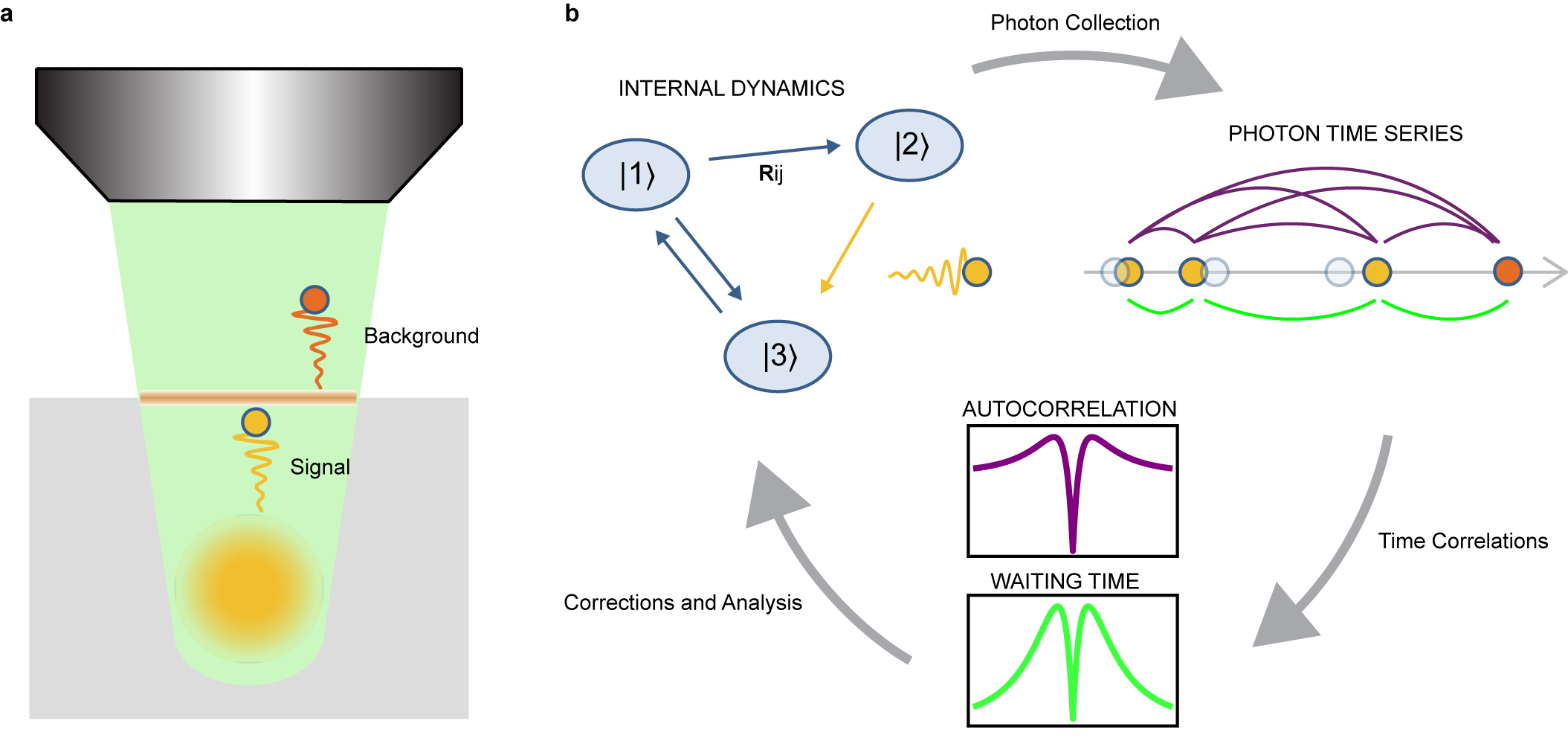}
\caption{Experimental overview. 
(a) Laser light is focused through a microscope objective onto a quantum defect in a solid-state crystal (grey block). 
The resulting fluorescence from the defect is emitted isotropically and collected through the objective as the signal. 
Background photons arising from surface fluorescence and other fluorescent defects are also collected. 
(b) The full process of PECS is illustrated. 
Starting in the upper left, excitation from a laser source causes the system to evolve between different electronic states, emitting a photon when passing through a radiative transition. 
The emitted photons (yellow circles) are collected into a photon time series, which includes experimental noise such as timing error, represented by light grey circles, and background photons (orange circle). 
Time correlations are calculated between either all photons or only subsequent photons to make up the autocorrelation or waiting time distribution respectively. 
Corrections and analysis of the photon emission statistics helps paint a clearer picture of the emitter's internal dynamics model.}
\label{fig:overview}
\end{figure*}


\section{INTRODUCTION}
Quantum defects originate from substitutional atoms, vacancies, or impurity-vacancy complexes in solid-state lattices. 
They can exhibit quantum-coherent spin and optical properties, and thus comprise foundational elements in quantum information science \cite{AharI2,Atat,Bass}.
Quantum defects are a subset of the larger category of quantum emitters, systems hosting discrete electronic states that interact with individual photons.
Examples include quantum dots, which are highly optimized single-photon sources for quantum photonics \cite{Hepp, Econ} and quantum communication \cite{AndeM,Arak}, fluorescent single molecules \cite{Bayl}, and solid-state point defects \cite{AharI2}. 
The latter category includes quantum defects that also exhibit spin coherence, which can couple to other classical and quantum degrees of freedom.
For example, fluorescent nanodiamonds containing quantum defects have advanced the field of quantum sensing \cite{Radt,Mill,Know}, and quantum registers in diamond are optically-interfaced quantum memories supporting multi-qubit quantum algorithms \cite{Child2,Brad,Cram,Abob}.
Inspired by the well-known examples of quantum dots \cite{Shan,Thom} and diamond color centers \cite{Dohe,Rodg}, the list of established quantum defect systems has grown to include defects in silicon carbide \cite{Bara, Soyk, Radu}, emitters in layered materials \cite{Pala} such as hexagonal boron nitride \cite{Toth} and transition metal dichalcogenides \cite{Liu}, and rare-earth ions \cite{Zhon}.
Most solid-state defect systems remain unexplored, however, and many promise potential advantages for quantum-information applications in terms of scalability, device integration, optical properties, spin properties, and quantum coherence \cite{Atat,Bass,Ferr}.
In each case, controlling and harnessing a defect's quantum properties requires a detailed understanding of its electronic structure as well as its optical and spin dynamics, presenting formidable obstacles for efficient experimental or theoretical characterization.

Photon emission correlation spectroscopy (PECS) is a valuable and often underutilized technique for elucidating a quantum emitter's optical and spin dynamics. 
PECS involves analyzing photon time correlations in the optical emission from a fluorescent system, as shown in Fig.~\ref{fig:overview}. 
It is widely used to verify single-photon emission associated with quantum emitters through the observation of photon antibunching \cite{Kimb7,Brou8,Cast,Mich,Tran,Kope}. 
As a steady-state measurement requiring only constant excitation, single-photon detectors, and suitable timing electronics, PECS is relatively simple to implement, and yet it can provide a wealth of information about an emitter's optical dynamics including excited-state lifetimes, radiative and non-radiative relaxation pathways, as well as spin and charge dynamics.  

This paper describes the application of PECS as a general-purpose characterization tool for solid-state quantum emitters. 
We present application-specific guidelines for reliable data acquisition, analysis, and interpretation.
In particular, we demonstrate how PECS can be used to reliably confirm single-photon emission and to hypothesize a model of the system’s electronic states and optical dynamics, enabling an assessment of the emitter’s suitability for quantum technology applications.



\section{BACKGROUND}
The method of optical intensity correlation spectroscopy was pioneered by Hanbury Brown and Twiss in 1956, when they recorded time-correlated photons while developing a new intensity interferometry technique to measure the diameter of stars, marking the first observation of the bunching of thermal light \cite{Hanb2}.
Glauber's foundational 1963 paper \cite{Glau}, laying the theoretical framework for higher order quantum correlation functions, launched the field of quantum optics.
Kimble \textit{et al.}'s 1977 observation of photon antibunching in emission from trapped ions confirmed the quantum-mechanical nature of light \cite{Kimb7,Wall}, and numerous subsequent experiments revealed phenomena including quantum jumps \cite{Berg11} and non-classical light fields \cite{Died9}.

The technique also found purchase in single-molecule spectroscopy \cite{Kurt6}, where innovations in microscopy had unveiled a new realm of molecular physics \cite{Grub5}. 
PECS facilitated exploration of intra- and intermolecular dynamics \cite {Gopi} that had previously been unresolvable for systems with fast timescales or low quantum yields \cite{Basc1}. 
Arguably, the largest contribution of PECS to single molecule spectroscopy was the development of Fluorescence Correlation Spectroscopy (FCS), which is widely used to resolve physical and kinetic dynamics such as diffusion rates, molecule size and orientation, blinking, and binding kinetics \cite{Mich, Elso}.

The emergence of quantum information science drove the application of PECS beyond single molecules to solid-state systems such as the nitrogen-vacancy (NV) center in diamond \cite{Kurt6,Brou8} and quantum dots \cite{Mich13}.
These systems promised potential platforms for single-photon sources that were more robust to photo-bleaching than molecules. 
The application of PECS to solid-state systems has since grown to include many other materials as a part of the search for optimal platforms for quantum technologies.

\subsection{\label{sec:ACfunction}The autocorrelation function\protect}

The primary way of characterizing photon correlations is through the second-order intensity correlation function, often called the autocorrelation function. In its most general form, the autocorrelation function is given by
\begin{equation}\label{eq:g2}
    g^{(2)}(\tau) = \frac{\langle I(t)I(t+\tau)\rangle}{\langle I(t)\rangle^{2}},
\end{equation}
where $I(t)$ is the intensity at time $t$, $\tau$ is the time delay between two intensity measurements, and $\langle \rangle$ represents the time-average of the enclosed quantity \cite{Loud3}. 

There are two important types of correlations that can appear in a measurement of $g^{(2)}(\tau)$.
Regions where $g^{(2)}(\tau)<1$, indicating a decreased probability of detecting two photons separated by $\tau$, are referred to as antibunching and correspond to a sub-Poissonian photon distribution.
Regions where $g^{(2)}(\tau)>1$, indicating increased detection probability, are referred to as bunching and correspond to a super-Poissonian photon distribution.
Any region where $g^{(2)}(\tau)=1$ corresponds to uncorrelated, Poissonian light.



\subsection{\label{sec:SPEs}Single-photon emitters\protect}

A single-photon emitter (SPE) is a quantum system that emits one photon at a time.
Single photons are a key requirement for many quantum information technologies \cite{Sene,Sinh}.
In particular, single-photon purity, the extent to which a system creates a pure, single-photon number state, influences the security of quantum communication protocols \cite{Leif,Take} and error rates in photonic quantum computing and simulation \cite{Sene}.
High purity single-photon emission is also a prerequisite for realizing indistinguishable single photons \cite{Mori}, which form the basis for linear-photonic quantum information processing protocols \cite{Knil,Scho,He} or quantum repeaters \cite{Pomp}.
Single-photon emission manifests in PECS measurements as an antibunching dip at zero delay.
Characterizing antibunching through PECS enables precise measurements of photon purity for SPEs \cite{Sene}



\subsection{\label{sec:optical_dynamics}Optical dynamics of quantum defects\protect}

The potential of quantum defects extends far beyond their use as SPEs.
When a defect's electronic or optical dynamics depend on internal orbital and spin states, these states become accessible as matter qubits for use in the storage or processing of quantum information.
The challenge in exploring new materials and defect systems is that a plethora of dynamical phenomena, including radiative and non-radiative transitions between electronic levels, spin dynamics, intersystem crossings to metastable states, and ionization/recombination charge transitions, may occur under different conditions, or all in combination.
Signatures of these phenomena manifest in the bunching dynamics of PECS measurements. As a result, PECS presents a versatile framework in which to hypothesize and test dynamical models.
In this section, we briefly discuss these phenomena and their importance for quantum information science.

Spin states are desirable as quantum-mechanical degrees of freedom because they are insulated from most environmental noise yet manipulable through spin resonance techniques, striking a balance between control and coherence \cite{Chat}.
When spin states couple coherently to light, they form a light-matter interface that is integral for quantum communication and distributed quantum computing as an interface between static and flying qubits \cite{Nort, Bassett2020}.
Even when the optical coupling to spin states is incoherent, as in the case of the NV center's intersystem crossing between triplet and singlet states, spin-dependent optical dynamics can be used for spin initialization \cite{Widm} and readout \cite{Hopp2}.  
Quantum sensing similarly takes advantage of the intrinsic sensitivity of orbital and spin states to external fields, together with optical readout \cite{Dege,Tayl}.
Many defects feature distinct spin manifolds separated by electric-dipole forbidden transitions. 
Forbidden transitions to shelving states with long lifetimes can allow a state to be stored and protected in quantum memories \cite{Hesh}.

Especially in wide-bandgap host materials, defects can exist in multiple stable charge states.
Once a defect's charge dynamics are understood and can be controlled, they present new opportunities for optical and electrical control, including electrical generation of single photons \cite{Schu} and long-term information storage \cite{Dhom}.
Charge states coupled to spin states can also be harnessed to significantly improve the efficiency of state initialization \cite{Hopp3} and optical readout \cite{Shie} for quantum computing or quantum sensing, or to enable photoelectric spin readout in microelectronic devices \cite{Siyu}.

More generally, detailed understanding of an emitter's electronic structure, along with radiative and non-radiative dynamics can also allow the design of additional resonant excitation schemes that improve spin readout efficiency \cite{Robl} or  achieve higher photon indistinguishability and entanglement \cite{Hube}.

\section{\label{sec:Theory}THEORY\protect}
PECS involves exciting and collecting emission from quantum emitters, often using a confocal microscope, as shown in Fig.~\ref{fig:overview}(a).
The experimental situation is similar for different types of emitters, including quantum dots, single molecules,and quantum defects, since these are all much smaller than the optical diffraction limit.
Figure~\ref{fig:overview}(b) presents an overview of the acquisition and analysis of PECS data.
The process begins with the internal dynamics of the quantum emitter system, which we assume is initially unknown.
The evolution of this unknown system in response to excitation determines the timing of photon emission.
The goal of PECS analysis is to infer the optical dynamics from the experimental data, and ultimately to develop a theoretical model for the quantum system.

\subsection{Types of photon correlations}

\begin{figure}[h!]
\includegraphics[trim = {0 0 0 0},width = \linewidth]{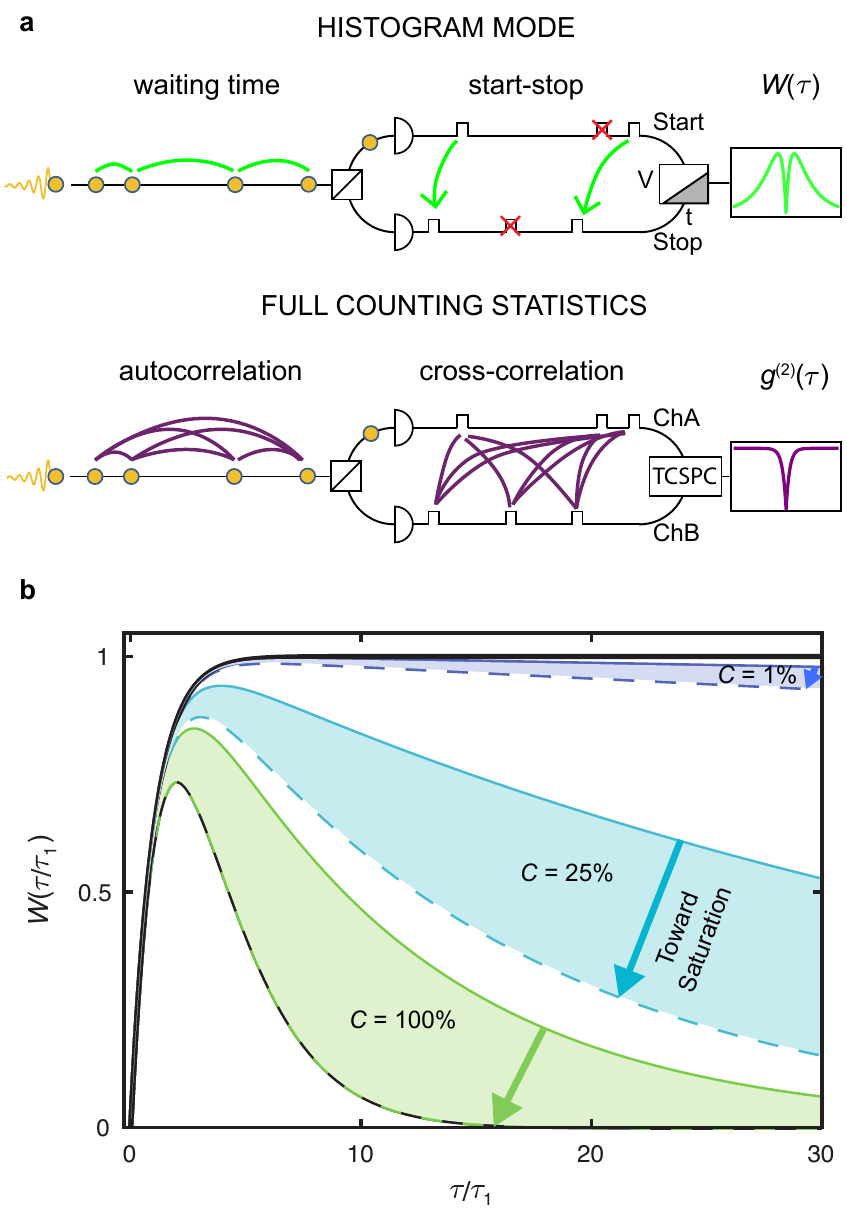}
\caption{\label{Figure2} Histogram mode and full counting statistics comparison.
(a) Schematic depicting the experimental setup for measuring histogram mode and the full counting statistics.
A snapshot in time shows photons and electric pulses travelling left to right.
A beamsplitter directs the photon stream into two separate detectors (semi-circles), each of which corresponds to a channel.
In histogram mode, one channel acts as a start pulse for a ramp circuit and the other is a stop pulse.
Additional start (stop) pulses that occur before (after) the stop (start) pulse are neglected.
For the full counting statistics, a time correlated single-photon counter (TCSPC) tags the absolute times of photon arrivals at each channel.
(b) The effect of collection efficiency and pump rate on the shape of the waiting time distribution for a two-level system. Collection efficiency is represented by differently-colored shaded regions, and pump rate relative to emission rate is represented on a spectrum from solid [maximally different pump and emission rates ($\alpha\rightarrow1$)] to dashed [pump rate = emission rate ($\alpha=0$)] lines.
The solid black curve represents $g^{(2)}(\tau/\tau_{1})$.
The dashed black curve represents the waiting time distribution at unity collection efficiency and equal pump and emission rates, at which point it is the farthest from approximating $g^{(2)}(\tau)$.
Traces have been normalized by pump rate and collection efficiency for ease of comparison.}
\label{fig:waitingTime}
\end{figure}

There are two types of photon intensity correlations that are often measured in experiments (see Fig.~\ref{fig:waitingTime}(a)). 
In addition to the autocorrelation function (Eq.~\ref{eq:g2}), which represents the likelihood of receiving any two photons separated by a specific time delay, the waiting time distribution, $W(\tau)$, depends only on correlations between subsequent photons.
Intuitively, $W(\tau)\mathrm{d}\tau$ represents the probability of detecting two subsequent photons with a time delay between $\tau$ and $\tau +\mathrm{d}\tau$.
Hence, $W(\tau)$ depends both on the dynamics of the system of study and on details of the experimental setup, such as the collection efficiency.
On the other hand, $g^{(2)}(\tau)$ captures correlations between all photons in the time series and reflects the full counting statistics of the system alone, independent of the collection efficiency. 

Figure~\ref{fig:waitingTime}(a) depicts the  experimental setups used to acquire $W(\tau)$ and $g^{(2)}(\tau)$.
The optical excitation is the same for both cases as is the use of a beamsplitter in the collection path to address detector dead time (see Sec.~\ref{sec:ExpConsid}).
However, the manner in which the collected photons are processed differs.
For $W(\tau)$, an incoming photon is registered as a start pulse, starting the clock until a subsequent photon is registered as a stop pulse.
This time difference is then collected into a histogram of photon time delays.
As a result, the method of acquiring $W(\tau)$ is referred to as histogram mode.
On the other hand, for $g^{(2)}(\tau)$, the arrival time of each photon is recorded, which requires a multi-channel, high-timing-resolution machine, such as a time-correlated single-photon counter, as well as additional processing to yield the correlations.
While $W(\tau)$ is often simpler to acquire experimentally, $g^{(2)}(\tau)$ is more straightforward to analyze for meaningful results. 

\subsection{Two-level model\protect}
The internal evolution of an emitter's states is determined solely based on the initial conditions, electronic states, and transition rates between the states.
Emission of a photon is dependent on which transitions are radiative.
Therefore, an analytic expression that captures the photon time correlations must be a function of the state of the system over time and must reflect which transitions are radiative.
The simplest model to consider is a two-level model, consisting of an excited state and a ground state.
The system transitions from the ground state to excited state at a rate $\Gamma_{ge}$ dependent on the excitation source, and then decays through a radiative transition from excited to ground at an intrinsic rate of $\Gamma_{eg}$, emitting one photon each time it decays.
Initially, we assume unity collection efficiency ($C$=1) in which every transition from the excited state to the ground state corresponds to a detected photon.
We will subsequently relax that assumption.

To derive an expression for $W(\tau)$ we must consider the probability of receiving the first subsequent photon at time $t_{2}$, given that a photon was received at time $t_{1}$.
For a two-level model, this is equivalent to the probability of the system starting in the ground state at time $t_{1}$, then evolving to the excited state at time $t$ after delay $\tau'=t-t_{1}$ and decaying back to the ground state at delay $\tau=t_{2}-t_{1}$, integrated over all possible excitation times:
\begin{equation}
    W(\tau) = \int_{0}^{\tau}d\tau'P_{e\rightarrow g}(\tau-\tau')P_{g\rightarrow e}(\tau').
    \label{eq:waittime}
\end{equation}
Here,
\begin{equation}
    P_{a\rightarrow b}(t) = \Gamma_{ab}e^{-\Gamma_{ab}t},
    \label{eq:pdef}
\end{equation}
is the normalized probability density function for a transition from state $|a\rangle$ to $|b\rangle$ with transition rate $\Gamma_{ab}$.
Therefore, for a two-level system with unity collection efficiency, the waiting time distribution is given by
\begin{equation}
    W(\tau) = \frac{\Gamma_{ge}\Gamma_{eg}}{\Gamma_{ge}-\Gamma_{eg}}(e^{\Gamma_{eg} \tau}-e^{-\Gamma_{ge} \tau}).
    \label{eq:waittime2lvl}
\end{equation}

To derive an equivalent expression for $g^{(2)}(\tau)$, we must consider the probability of receiving \emph{any} photon at time $t_{2}$, given one was received at time $t_{1}$.
For any model with a single radiative transition, this is equivalent to $P_{e}(t_{2}|P_{g}(t_{1})=1)$, the probability of being in the excited state at time $t_{2}$, given that the system was in the ground state at $t_{1}$.
Normalized by the steady-state population of the excited state, $P_{e}^{\infty}$, this gives the autocorrelation function,
\begin{equation}
    g^{(2)}(\tau=t_{2}-t_{1}) = \frac{P_{e}(t_{2}|P_{g}(t_{1})=1)}{P_{e}^{\infty}},
    \label{eq:autocorrelation}
\end{equation}
so that $g^{(2)}(\tau)=1$ corresponds to uncorrelated light and any deviations from 1 correspond to positive or negative correlations.
 
The time-dependent probability of each state's occupation is determined by the transition rates.
Therefore, the probability of excited state occupation can be found by solving a system of coupled ordinary differential equations (ODEs), which for a two-level model is
\begin{subequations}
\begin{align}
    \frac{dP_{g}}{dt} &= -\Gamma_{ge}P_{g}(t)+\Gamma_{eg}P_{e}(t) \\
    \frac{dP_{e}}{dt} &= \Gamma_{ge}P_{g}(t)-\Gamma_{eg}P_{e}(t).
\end{align}
\end{subequations}
Solving with the initial conditions of 
\begin{equation}
    P_{g}(0)=1, 
    \label{eq:initCondit}
\end{equation}
results in the expression for autocorrelation from a two-level model,
\begin{equation}
    g^{(2)}(\tau) = 1-e^{-(\Gamma_{eg}+\Gamma_{ge})\tau}.
\end{equation}

\subsection{Generalizing to models with $n>2$ levels\protect}
In order to capture more complicated dynamics, models with $n>2$ levels are necessary.
Experimentally observed optical dynamics can often be the product of multiple electronic levels involving additional radiative and non-radiative transitions.

One example of a multi-level model, a three-level model, might include a two-level model with an additional non-radiative pathway from the excited state to a third metastable state, then to the ground state.
In the following, we will consider the general cases of multi-level models with a single radiative transition and unity collection efficiency.

In the case of $W(\tau)$, the addition of non-radiative decay pathways requires accounting for all possible combinations of non-radiative loops through electronic states that can occur before the emission of a second photon after delay $\tau$. For a general system with $n$ possible non-radiative decay pathways from the excited state back to the ground state, 

\begin{equation}
    W(\tau) = P_{g\rightarrow e}(\tau)*P_{e\rightarrow g}(\tau)*\left(1+\sum_{k=1}^{\infty}\left[\left(\sum_{i=1}^{n}h_{i}(\tau)\right)^{*k}\right] \right).
    \label{eq:WaitGeneral}
\end{equation}
Here $h_{i}(\tau)$ is the probability density function for evolution through each non-radiative decay loop, $i$, starting and ending in the ground state with a total duration $\tau$, and
\begin{equation}
    h(\tau)^{*n} = \underbrace{h(\tau)*h(\tau)*...*h(\tau)}_{n}.
\end{equation}
The prefactor of Eq.~\ref{eq:WaitGeneral} represents the PDF of travelling through the radiative loop one time, while the part enclosed in parentheses represents all possible combinations of non-radiative loops.
With increasingly complex models, evaluating $W(\tau)$ quickly becomes intractable.

Generalizing $g^{(2)}(\tau)$ to multi-level models is more straightforward.
As in the case of the two-level model, for any $n$-level electronic structure, the full dynamics are given by a system of $n$ coupled differential equations.
This system of equations can be summarized by the rate equation
\begin{equation}
    \dot{P} = GP,
    \label{eq:master}
\end{equation}
where $P$ is a vector of state occupation probabilities and $G$ is the transition rate matrix. 
Each off-diagonal element of the rate matrix, $G_{ij}$, where $i\neq j$, is the total transition rate into state $|i\rangle$ from state $|j\rangle$. 
%
Each diagonal element $G_{ii}=-\sum_{j\neq i}G_{ji}$ is the total transition rate out of state $i$ and thus preserves probability.
The time-dependent population of each state can be obtained by solving Eq.~\ref{eq:master} with the initial condition set immediately following emission of a photon (Eq.~\ref{eq:initCondit} for systems with a single ground state, and Eq.~\ref{eq:P0} for multiple ground states).
$g^{(2)}(\tau)$ can then be calculated from Eq.~\ref{eq:autocorrelation}, letting $P_{e}$ be the population of the excited state from which the radiative transition occurs. 
Additional derivations of the waiting time distribution as a function of collection efficiency and for mutli-level models can be found in Appendix~\ref{sec:apxWait}.
Discussion of autocorrelation for models with multiple radiative transitions can be found in Appendix ~\ref{sec:apxMultRad}.

\subsection{Relationship between $W(\tau)$ and $g^{(2)}(\tau)$\protect}
$W(\tau)$ and $g^{(2)}(\tau)$ can be analytically related, as both sets of correlations originate from the same physical process, and the correlations contained in $W(\tau)$ make up a subset of all those included in $g^{(2)}(\tau)$.
As a result, $g^{(2)}(\tau)$ can be constructed from $W(\tau)$ through an infinite sum of self-convolutions \cite{Cohe2}
\begin{equation}
\begin{split}
    g^{(2)}(\tau)& = W(\tau)+W(\tau)*W(\tau)+\cdots \\
    &= \mathcal{L}^{-1}\left\{
    \frac{\mathcal{L}\{W\}(s)}{1-\mathcal{L}\{W\}(s)}\right\}(t),
    \label{eq:g2WtConv}
\end{split}
\end{equation}
where $\mathcal{L}$ is the Laplace transform and $s$ is a complex frequency parameter.
This is due to the fact that the probability of receiving two photons separated by time $\tau$ with $m$ intermediate detection events is equivalent to $m$ convolutions of the probability of receiving consecutive photons.


Equation~\ref{eq:g2WtConv} shows how $W(\tau)$ can be thought of as a first order approximation of $g^{(2)}(\tau)$.
This relationship has led to the occasional practice of using $W(\tau)$ and $g^{(2)}(\tau)$ interchangeably in experiments.
However, $W(\tau)$ depends dramatically on the apparent brightness of the signal (see Appendix~\ref{sec:apxWait}), while $g^{(2)}(\tau)$ does not.
Therefore, the accuracy of this approximation is tied to the setup collection efficiency, $C$, and $\alpha$, the relation between the pump rate ($\Gamma_{ge}$) and radiative decay rate ($\Gamma_{eg})$,
\begin{equation}
   \alpha = \frac{\abs{\Gamma_{eg}-\Gamma_{ge}}}{\Gamma_{eg}+\Gamma_{ge}},
\end{equation}
both of which impact apparent brightness.

Fig.~\ref{fig:waitingTime}(b) illustrates the effect of $C$ and $\alpha$ on the shape of $W(\tau)$ and its comparison to $g^{(2)}(\tau)$ (solid black line) for a two-level model.
Each shaded region depicts a set of $W(\tau)$ curves at a particular $C$, as a function of $\alpha$.
Within a shaded region, the shape of $W(\tau)$ ranges from a solid colored line, representing $W(\tau)$ for a system in which the pump rate is significantly higher or significantly lower than the decay rate ($\alpha\rightarrow1$), to a dashed line, representing a system in which the two rates are equal ($\alpha=0$).
A lower collection efficiency generally leads to a closer approximation of $g^{(2)}(\tau)$.
On the other hand, adjusting the excitation power so that the pump rate approaches the emission rate causes $W(\tau)$ to diverge from $g^{(2)}(\tau)$.
The difficulty in determining the experimental quantities of $C$ and $\alpha$ make it challenging to assess the validity of approximating $g^{(2)}(\tau)$ with $W(\tau$) in practical situations.
Further, the dependence of $W(\tau)$ on these three independent variables make it difficult to decouple collection efficiency from transition rates when measuring an unknown system.
For this reason, it is almost always preferable to acquire $g^{(2)}(\tau)$ for quantitative analysis.
Therefore, the remainder of this text will focus on the use of $g^{(2)}(\tau)$.

\subsection{Single-photon emission\protect}
PECS is an ideal measurement to characterize single-photon emission. 
An ideal SPE can only emit one photon at a time, and hence the probability to observe two photons with zero delay, and correspondingly $g^{(2)}(0)$, must equal zero.
This fact is typically justified by quantum optics arguments.  
For a photon number state $\left |n\right\rangle$ with exactly $n$ photons, 
\begin{equation}
    g^{(2)}(0) = \frac{\langle\hat{n}(\hat{n}-1)\rangle}{\langle\hat{n}\rangle^{2}} = \frac{(n-1)}{n}.
\label{eq:SinglePhotonCriteria}
\end{equation}
Using this relationship, it is apparent that $g^{(2)}(0)=0$ for $n=1$ and that $g^{(2)}(0)\geq0.5$ for $n\geq 2$.
Hence, it has become common practice to check whether an emitter's measured $g^{(2)}(0)$ is less than $0.5$.

However, the use of the so-called  ``0.5 criterion'' is questionable, since it does not accurately reflect the situation encountered in most experiments with quantum emitters.
As derived, Eq.~\ref{eq:SinglePhotonCriteria} applies to photon number states, which only occur if photons are emitted by identical, two-level emitters into the same spatial and temporal modes \cite{Loud3}.
In typical experiments, however, uncorrelated emission from $n$ independent, nonidentical emitters does not create photon number states.
Hence, the criteria for establishing single-photon emission needs to be re-evaluated.


Generalizing Eq.~\ref{eq:autocorrelation}, the autocorrelation function measured from $n$ emitters is proportional to the sum of a correlated probability that two photons are received from the same emitter and an uncorrelated probability that two photons are received from different emitters.
In the case where emitter $i$ has brightness $I_{i}$ this gives
\begin{equation}
\begin{split}
    &g^{(2)}(\tau;n) = \\ &\sum_{i=1}^{n}\frac{I_{i}}{\sum_{k=1}^{n}I_{k}}\left(\frac{(P_{e}^{i}(t_{2}|P_{g}^{i}(t_{1})=1)}{P_{e}^{i,\infty}}  +\sum_{j\neq i}^{n}\frac{I_{j}}{\sum_{k=1}^{n}I_{k}}\right).
    \label{eq:multEmitters_2}
\end{split}
\end{equation}
As $\tau$ approaches 0 (as $t_{2}$ approaches $t_{1}$), the first term goes to zero, giving the normalized expression for $g^{(2)}(0)$ from multiple emitters:
\begin{equation}
    g^{(2)}(0;n) = \frac{\sum_{i=1}^{n}\sum_{j\neq i}^{n}I_{i}I_{j}}{(\sum_{k=1}^{n}I_{k})^{2}} = \frac{(\sum_{k=1}^{n}I_{k})^{2}-\sum_{k=1}^{n}I_{k}^{2}}{(\sum_{k=1}^{n}I_{k})^{2}}.
    \label{eq:multEmitters_t0}
\end{equation}
For $n$ emitters with identical brightness, $I_j=I \forall j$, this derivation returns Eq.~\ref{eq:SinglePhotonCriteria}.
However, Eq.~\ref{eq:SinglePhotonCriteria} \emph{only holds} in the case of emitters with identical brightness.
For example, in the case where $n$=2, Eq.~\ref{eq:multEmitters_t0} reduces to
\begin{equation}
    g^{(2)}(0;2) = \frac{2I_{1}I_{2}}{(I_{1}+I_{2})^{2}}.
\end{equation}
If $I_{2}>I_{1}$ such that $I_{2} = I_{1}+\delta$, we find
\begin{equation}
    g^{(2)}(0;2) =  \frac{1}{2+\frac{\delta^{2}}{2I_{1}(I_{1}+\delta)}}.
\end{equation}
The second term in the denominator is always positive, and hence  $g^{(2)}(0)<0.5$.

Quantum emitters are typically not identical. 
Even when they are the same species, a variety of factors, including proximity to surfaces and alignment of the excitation or emission dipoles, can influence their observed brightness.
Hence, the $g^{(2)}(0)<0.5$ criterion is insufficient to identify single-photon emitters.
It can be erroneously satisfied even when multiple emitters are present.

In contrast, a measurement of the ideal relationship $g^{(2)}(0)=0$ would confirm single-photon emission.
In order to apply this stricter criterion to experiments, it is necessary to account for systematic and stochastic errors that can lead to measurements of $g^{(2)}(0)>0$ even for a SPE.
The next section shows how to account for these effects, in order to achieve measurements of $g^{(2)}(0)=0$ within quantified uncertainties for a SPE.

\section{\label{sec:ExpConsid}EXPERIMENTAL CONSIDERATIONS\protect}

Experimental acquisition and analysis of photon emission statistics present a number of challenges that must be considered in conjunction with the idealized theory from the previous section.
A proper experiment involves processing significant amounts of data, and one must account for the timing resolution of detectors and correct for systematic experimental artifacts.
Here we discuss the experimental setup for the collection of photon emission statistics, highlight an efficient algorithm to aid in calculating $g^{(2)}(\tau)$ from photon arrival times, and describe how to correct for the dominant sources of experimental error, \textit{i.e.}, background photons that did not come from the emitter and detector timing jitter.




\subsection{Acquisition\protect}
Photon emission statistics measurements of quantum emitters are typically acquired using a confocal microscope.
In contrast to a wide-field microscope, a confocal arrangement rejects background emission from regions of the sample outside a diffraction-limited volume around the emitter of interest, and it directs the collected photons to a single detector channel that can be optimized for detection efficiency and timing resolution.
Most detectors suffer from dead time, which is a period following each photon detection event during which the detector is blind to subsequent photons.
In order to measure dynamics within the detector dead time, a beamsplitter is introduced in the emission path, directing the photon stream into two different, but nearly identical, detectors (see Fig.~\ref{fig:waitingTime}(a)).
In this way, the autocorrelation function of the original photon stream is directly related to the cross-correlation function calculated  across the two detectors, each corresponding to a channel.
While $W(\tau)$ can be acquired by measuring only relative times through start-stop collection, $g^{(2)}(\tau)$ requires time tagging each photon detection event and subsequently calculating the photon correlations across the two channels.

\subsection{Data processing\protect \label{sec:DataProc}}
A brute-force calculation of the cross correlation function for large set of time-tagged data is extremely time consuming even for modern computers.
Fortunately, Laurence \textit{et al.} described a more efficient algorithm  \cite{Laur}.
Rather than individually iterating through all photon pairs between channels and binning the results into the photon correlation function, Laurence \textit{et al.}'s algorithm uses the fact that the data are sorted in time to substantially reduce the processing time.
Their method also allows for arbitrarily defined bins that are not equally sized.
In interpreting PECS data, it is often useful to utilize logarithmically-varying time bins, in order to visualize and analyze correlations occurring at widely-varying time scales.

We have incorporated the algorithm developed by Laurence \textit{et al.} into a library of MATLAB functions for calculating and visualizing the autocorrelation function obtained from raw PECS data.
In addition to $g^{(2)}(\tau)$, the library functions also calculate the time-averaged, steady-state intensity  over the course of acquisition, which can be referenced to control for experimental factors such as fluctuations in emitter stability and setup drift.
Blinking, in particular, is a common problem for quantum emitters.
Blinking emitters stochastically switch between two or more brightness levels due to fluctuations in their local environment.
Our PECS analysis code allows for the data to be thresholded according to the time-averaged intensity, with the autocorrelation function calculated separately for different brightness levels.
This can enable detailed studies of optical dynamics even for stochastically blinking emitters.
Setup drift can typically be reduced by implementing a tracking scheme that periodically adjusts the microscope alignment between successive autocorrelation measurements.

Our implementation of  Laurence \textit{et al.}'s algorithm also returns the statistical uncertainty of a PECS measurement.
The algorithm iterates through photons in Channel A, calculating correlations to bins in Channel B.
Uncertainty in PECS data are dominated by shot noise; hence, if the number of photons recorded in a given bin is $M$, the Poissonian uncertainty is $\sqrt{M}$.
To calculate $g^{(2)}(\tau)$, each bin is normalized by its time-averaged, expected number of counts, $I_A I_B T w$, where $I_A$ and $I_B$ are the time-averaged count rates in each channel, $T$ is the total acquisition time, and $w$ is the bin width.
Hence, the experimental uncertainty in $g^{(2)}(\tau)$ is
\begin{equation}
\Delta = \frac{\sqrt{M}}{I_\textrm{A}I_\textrm{B}wT}.
\label{eq:poissonErr}
\end{equation}
For additional detail about the processing algorithm and a discussion of asymmetric errors, see Appendix~\ref{sec:apxAlg}.

\subsection{Correcting for background signals\protect}
Once the correlations have been processed, the next step involves correcting the data for background.
Background photons can arise from dark counts of the detection system, fluorescence of the host material, or other sources of room light.
These background signals result in an inflated likelihood of observing uncorrelated light.
As a result, background signals compress the $g^{(2)}(\tau)$ function toward 1, decreasing the extent of deviation above or below 1 at all delays.

The effect of a background signal with average intensity $I_\textrm{bg}$ on $g^{(2)}(\tau)$ for an emitter of intensity $I_\textrm{em}$ can be derived following a similar logic to the derivation of Eq.~\ref{eq:multEmitters_2}. 
This results in the background incorporated expression \cite{Brou8},
\begin{equation}
    g^{(2)}_\textrm{meas}(\tau) = 1-\rho^{2}+g^{(2)}_\textrm{}(\tau)\rho^{2},
    \label{eq:g2raw}
\end{equation}
where $\rho = \frac{I_\textrm{em}}{I_\textrm{em}+I_\textrm{bg}}$.
As a result, correcting for background only requires measuring $\rho$
This can be achieved in various ways, including measuring $I_{\textrm{bg}}$ at a point outside the diffraction-limited volume around the emitter, fitting the emitter's spatial profile to obtain $I_{\textrm{bg}}$ and $I_{\textrm{em}}$, or fitting excitation-power-dependent photoluminescence intensity data using a known saturation function for the emitter and a linearly-scaling background component.

\begin{figure}
\includegraphics[trim={0 0 0 0},width = \linewidth]{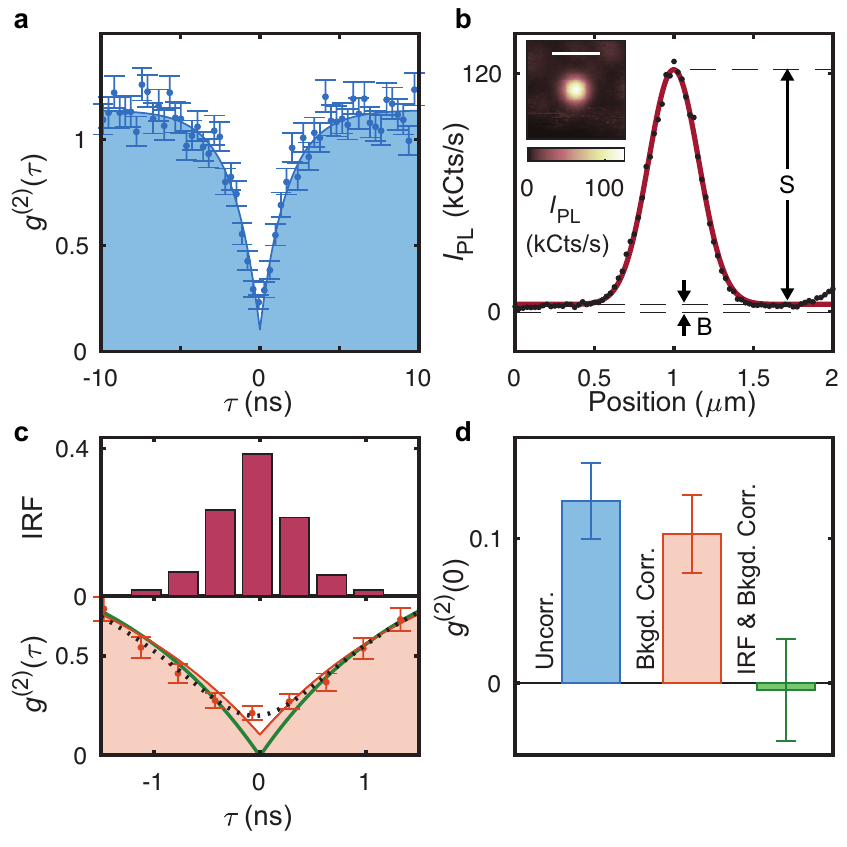}
\caption{Background and timing jitter correction to verify single-photon emission in hexagonal boron nitride (hBN).
(a) Raw photon emission statistics data from an emitter in hBN.
(b) Transverse intensity trace across the $y$ cross-section of the PL scan in inset taken at the focus plane.
White scale bar in inset shows 1$\mu$m.
Signal and background are denoted by arrows and are extracted from fit.
(c) (lower panel) Background-corrected data and fit (orange), and extracted $g^{(2)}(\tau)$ after background and timing-jitter correction (green).
Convolution of the fully corrected data with the measured instrument response function (upper panel) gives the black dotted line.
(d) The value of $g^{(2)}(0)$ from the fit before corrections (blue), after background correction only (orange), and after background and timing jitter correction (green). Uncertainties in $g^{(2)}(0)$ are 68\% confidence intervals propagated from the corresponding best fits.}
\label{fig:Corrections}
\end{figure}

Figure~\ref{fig:Corrections}(a) shows an example of raw  $g^{(2)}(\tau)$ data acquired from an emitter in room-temperature, hexagonal boron nitride (hBN) prior to any corrections.
Figure~\ref{fig:Corrections}(b) illustrates the process of fitting an intensity line trace in order to measure $\rho$ so that background correction can be performed.
Transverse intensity line traces across $x$ and $y$ cross-sections (data for $y$ cross-section are shown) of the emitter's 2D photoluminescence (PL) scan at the focal plane (inset) are fit using Gaussian functions, with the peak amplitude and offset of the fits giving the values for signal and background, respectively. 
The background is then corrected by solving Eq.~\ref{eq:g2raw} for $g^{(2)}_\textrm{}(\tau)$.
In Fig.~\ref{fig:Corrections}(c), the orange data points and orange shaded fit show the resulting $g^{(2)}(\tau)$ data and fit following background subtraction.

The emitter shown in Fig.~\ref{fig:Corrections} contributes 99\% of the total signal, so background photons have a minimal effect on the shape of $g^{(2)}(\tau)$ and the value of $g^{(2)}(0)$ (see panels C and D).
However, this is not always the case.
In many situations, background photons can be a dominant source of systematic error.
In these cases, it can also be helpful to quantify the excitation power dependence of the emission rate. 
Since the emitter signal typically saturates with increased excitation power, whereas background signals scale linearly, one can select an excitation condition where the signal-to-background ratio is maximized. 
Most emitters are characterized by saturation functions that approximate the empirical form \cite{Bech}
\begin{equation}
    I(p) = \frac{\Isat p}{p_{\textrm{sat}}+p}+C_{\textrm{bg}}p,
\end{equation}
where $p$ is the excitation power, $I_{sat}$ is the saturation intensity, $p_{sat}$ is saturation power, and $C_{bg}$ is the contribution from the background. 
Often, acquiring photon emission statistics close to saturation power balances the desire for high $\rho$ and sufficient signal to minimize shot noise.
In a two-level system, saturation power corresponds to the situation when the excitation rate equals the emission rate, $\Gamma_{eg}$ = $\Gamma_{ge}$.
However, one must also consider the fact that antibunching and bunching timescales are generally a function of excitation power, as described in Section~\ref{sec:Theory}.


\subsection{Correcting for timing jitter\protect}
Detector timing jitter, also known as the instrument response function (IRF), is the distribution of the electronic response time of the detector system to signal an event after photon arrival.
Integrating the IRF over a time range gives the probability of the detector registering a photon event within that time window after the photon is received,
The IRF of an ideal detector system is a delta function, but for a realistic experiment, the distribution will have a non-zero width.
While timing jitter can arise from any electronics in the system that add arrival time uncertainty, the choice of detector typically has the largest contribution to the IRF \cite{Wahl}.
Commonly-used single-photon avalanche diode detectors typically have IRF widths ranging from 100~ps to 1~ns.

The timing error manifests in the $g^{(2)}(\tau)$ trace as a convolution of the timing error distribution with the actual $g^{(2)}(\tau)$ signal from the emitter, \textit{i.e.},
\begin{equation}
g^{(2)}_{\textrm{meas}} = \textrm{IRF} * g^{(2)}.
\label{eq:IRFConv}
\end{equation} 
The convolution changes the measured value of $g^{(2)}(0)$ and the shape of $g^{(2)}(\tau)$ at small delays comparable to the IRF width.

Correcting for the timing jitter requires measuring the IRF of the setup.
The IRF can be obtained by collecting the distribution of detection times from a highly attenuated ($\sim$0.1 photons/pulse), pulsed laser source with a pulse width much less than the specified timing jitter of the detectors.
When using two detectors to measure photon emission statistics, the IRF of both detectors can be acquired by measuring the autocorrelation from the pulsed source.
This measurement will give a convolution of the two detectors' timing distributions and the shape of the pulsed source. 
However, when the optical pulse width is much less than the IRF width, it can be neglected.
While some IRFs can be approximated as Gaussian, the shape of the IRF can vary depending on the detector, and the functional form may not always be obvious \cite{Stev}.

Once the IRF of the setup is measured, the $g^{(2)}(\tau)$ data can be compensated for its systematic effects.
One method, deconvolution, involves solving Eq.~\ref{eq:IRFConv} for $g^{(2)}(\tau)$.
However, deconvolution amplifies noise and complicates propagation of experimental uncertainty.
Therefore, it is often preferable to incorporate the measured IRF into a fitting function to be compared directly with the measured $g^{(2)}(\tau)$ data.
This can be accomplished by including the numerical convolution of the measured IRF within the emprical fit function for $g^{(2)}(\tau)$.
This method requires that the measured IRF and $g^{(2)}(\tau)$ be processed with the same, uniform, time bin width.
The true timescales of the emitter and uncertainties can be extracted from the resulting best fit.

Figure~\ref{fig:Corrections}(c) illustrates an example of IRF correction.
The measured IRF data are shown in red (top) and are binned with a 350 ps bin width, as are the measured $g^{(2)}(\tau)$ data (orange, bottom), here shown after background correction but before IRF correction.
The green line displays the IRF-corrected $g^{(2)}(\tau)$ empirical best fit.
The black dotted curve, which is a convolution of the green line and IRF, aligns closely with the data and is used to determine best-fit parameters and uncertainties using a least-squares fitting method.


\subsection{Quantifying the effects of timing jitter}

The extent of the IRF's effect on the shape of $g^{(2)}(\tau)$ hinges on how its standard deviation width, $\sigma$, compares to the internal timescales of the emitter.
In the cases where $\sigma \gtrsim \tau_{1}$, where $\tau_{1}$ denotes the shortest timescale to emit subsequent photons (typically, the antibunching timescale), the emitter's faster dynamics can be obscured by the timing jitter.

\begin{figure}
\includegraphics[trim={0 0 0 0},width = \linewidth]{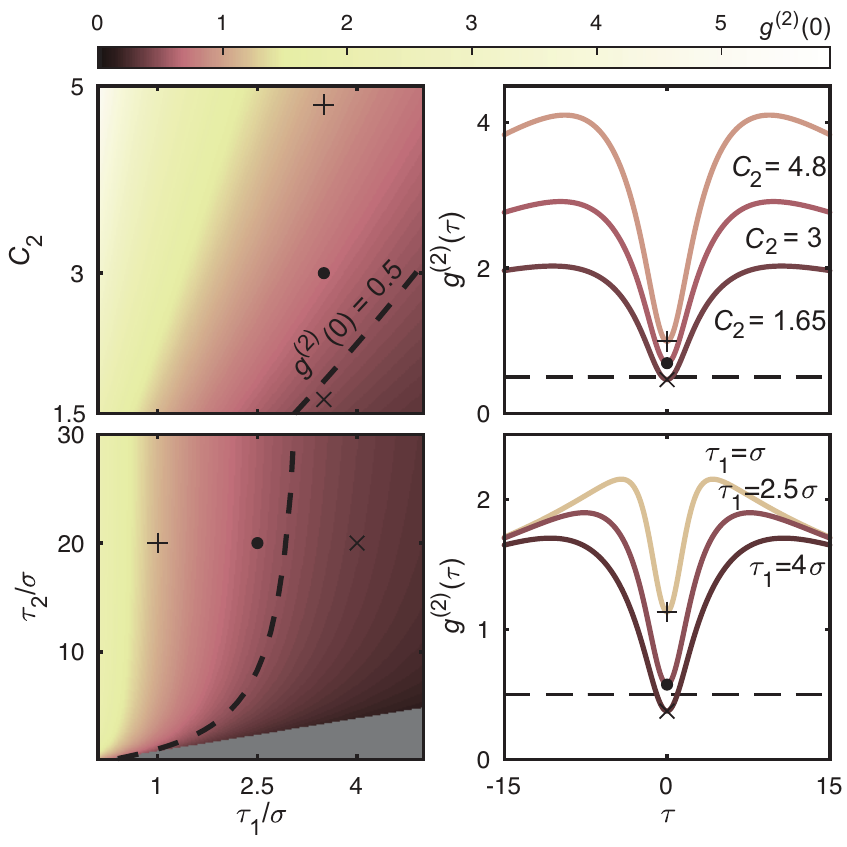}
\caption{\label{Figure4} Effect of timing jitter on $g^{(2)}(0)$.
(Left) The value of $g^{(2)}(0)$ is calculated as a function of the ratio of $\frac{\tau_{1}}{\sigma}$ and $\frac{\tau_{2}}{\sigma}$ or $C_{2}$ for a three-level system with timescales $\tau_{1}$ and $\tau_{2}$ and bunching amplitude $C_{2}$, and a gaussian IRF with standard deviation $\sigma$.
(Right) Six points are selected from the parameter combinations on the left to illustrate how $g^{(2)}(\tau)$ at low delays changes for different ratios of $\frac{\tau_{1}}{\sigma}$ (bottom) and different values of $C_{2}$ (top).}
\label{fig:IRF}
\end{figure}

The combination of bunching and antibunching dynamics on different timescales can further complicate the effects of the detector IRF.
Figure ~\ref{fig:IRF} illustrates an example of the effect of a Gaussian IRF with width $\sigma$ on $g^{(2)}(\tau)$ of an emitter represented by a three-level (two-timescale) model:
\begin{equation}
    g^{(2)}(\tau) = 1-(1+C_{2})e^{\frac{-\abs{\tau}}{\tau_{1}}}+C_{2}e^{\frac{-\abs{\tau}}{\tau_{2}}},
\end{equation}
with antibunching timescale $\tau_{1}$, bunching timescale $\tau_{2}$, and bunching amplitude $C_{2}$.
The left-hand side of the figure depicts the value of $g^{(2)}(0)$ as a function of different parameters, and the right-hand side depicts the effect on the whole shape of $g^{(2)}(\tau)$ for select combinations.
A dashed line shows the threshold for measuring $g^{(2)}(0)$ = 0.5. 
The upper two panels examine the effect for different bunching amplitudes at a fixed ratio of $\frac{\tau_{2}}{\sigma}=30$.
Hence in these cases, the bunching timescale is much larger than the IRF width.
Nonetheless, the higher the bunching amplitude, the greater effect the convolution of the IRF and $g^{(2)}(\tau)$ have on the measured $g^{(2)}(\tau)$ at low times, an effect which is amplified for low $\frac{\tau_{1}}{\sigma}$.
The lower two panels examine the effect of the ratio $\frac{\tau_{1}}{\sigma}$ for fixed $C_{2}=1.5$.
As $\tau_{1} \rightarrow \sigma$ from above, the measured value of $g^{(2)}(0)$ increases and the width of the antibunching dip at short delays decreases. For systems with more than three levels, we would expect similar effects, with the IRF impacting measurements of $g^{(2)}(0)$ and the shortest timescale, $\tau_{1}$, the most.
These effects exemplify how IRF correction can play a critical role in extracting the actual value of $g^{(2)}(0)$ and confirming single-photon emission. An example of such a case is illustrated by the hBN emitter in Fig.~\ref{fig:Corrections}(d).

\section{ANALYSIS}

The ability to analyze PECS data in order to infer an emitter's internal dynamics requires an understanding of how stochastic evolution through radiative and non-radiative states in an electronic model leads to features in $g^{(2)}(\tau)$.

The rate equation governing the population dynamics, Eq.~\ref{eq:master}, is a first-order, linear ODE, with general solutions of the form
\begin{equation}
    \vec{P}(t) = A_{0}\vec{v}_{0} +  \sum_{i=1}^{n-1}A_{i}e^{\lambda_{i}t}\vec{v}_{i}.
    \label{eq:generalSoln}
\end{equation}
Here, $\lambda_i$ are the eigenvalue rates, $v_i$ are the eigenvectors, and $A_i$ are constants determined by the initial condition.
For an $n$-level system with a single excited state, the excited state probability is
\begin{equation}
    P_{e}(t) = A_{0}(\vec{v}_{0}\cdot \hat{e}) + \sum_{i=1}^{n-1}A_{i}(\vec{v}_{i}\cdot \hat{e})e^{\lambda_{i}t}.
    \label{eq:empiricalPe}
\end{equation}

The probability-conserving condition of $\Sigma_{j}G_{ij}=0$ means that there will always be a zero eigenvalue, $\lambda_{0}$, and null eigenvector, $v_{0}$, corresponding to the solution of the steady-state equation,
\begin{equation}
    0 = GP.
    \label{eq:steadystate}
\end{equation}
The non-zero eigenvalues $\lambda_i$ can be real or complex, depending on the properties of $G$.  
If complex eigenvalues do appear, they occur in conjugate pairs due to the real, non-negative transition rates.
In this way, the general solution remains real.
In all cases, the real part of any non-zero eigenvalues will be negative \cite{Timm}. 
Therefore, from the eigenvalues, we can define a set of timescales governing different processes, $\tau_{i}=-\frac{1}{\textrm{Re}(\lambda_{i})}$ where $\tau_{i}>0$. 
Given an initial condition corresponding to the system configuration following the detection of a photon, we can follow Eq.~\ref{eq:autocorrelation} to obtain a general form of the autocorrelation function.
Normalizing Eq.~\ref{eq:empiricalPe} to the steady state, with the assumption of no background such that any detected photon projects the system into the ground state, this results in the following general empirical formula:
\begin{equation}
    g^{(2)}(\tau) = 1+\sum_{i=1}^{n-1}C_{i}e^{-\frac{\tau}{\tau_{i}}},
    \label{eq:empiricalFit}
\end{equation}
where $C_{i} = \frac{A_{i}(\vec{v}_{i}\cdot \hat{e})}{P_{e}^{\infty}}$ are constants, $P_{e}^{\infty}=A_{0}(\vec{v}_{0}\cdot \hat{e})$ is the steady-state excited state population, and $n$ is the number of states.
Equation~\ref{eq:empiricalFit} defines a curve that starts at $0$ for $\tau=0$ and decays to $g^{(2)}\rightarrow 1$ as $\tau\rightarrow \infty$. 

Antibunching 
arises when emission of consecutive photons is delayed as the excited state is re-populated, leading to a decreased likelihood of photons separated by short times.
Empirically, it is captured by terms in Eq.~\ref{eq:empiricalFit} with negative prefactors.
In the case of a two-level model, antibunching occurs on the timescale of $\tau_{1} = \frac{1}{\Gamma_{ge}+\Gamma_{eg}}$, representing the time to evolve from the ground state to the excited state and back to the ground state again.
Bunching dynamics, on the other hand, arise from transitions to non-radiative states, which delay the emission of a photon, such as transitions between charge or spin manifolds.
Such processes can result in the emitter's excited state population(s) evolving non-monotonically toward the steady state, leading to bunching in the autocorrelation trace.
When multiple non-radiative states participate in the dynamics, with different lifetimes, the autocorrelation function features multiple, resolvable bunching timescales.
In some situations, multiple radiative excited states can lead to multiple antibunching terms and complex eigenvalues associated with the antibunching dynamics \cite{Pate}.
However, such situations are uncommon and multiple antibunching rates are typically difficult to resolve experimentally. 
Therefore, it is typically appropriate to assume a single antibunching timescale, with a corresponding empirical model,
\begin{equation}\label{eq:empirical_g2}
    g^{(2)}(\tau) = 1-C_{1}e^{-\frac{\tau}{\tau_{1}}}+\sum_{i=2}^{n-1}C_{i}e^{-\frac{\tau}{\tau_{i}}},
\end{equation}
where all the $C_{i}$ are positive.
Given PECS data from an emitter with unknown level structure and dynamics, the set of models for varied $n$ can be fit to the data, and a statistical comparison based on the Akaiki Information Criterion or the chi-squared statistic can determine the most appropriate model to describe the data.
A determination of $n$ in this way places a lower limit on the number of  electronic levels involved in the dynamics. 
Additional details on fit comparisons using the Akaike Information Criterion can be found in Appendix~\ref{sec:apxAkai}.
Potential electronic models can be further narrowed down by measuring $g^{(2)}(\tau)$ at different powers and fields and comparing with simulations, as we describe in the next subsection.

\subsection{Simulating photon emission statistics}

Simulations of $g^{(2)}(\tau)$ provide a means to test potential models that explain features observed in experimental data.
The time-dependent state populations of a given model, consisting of $n$ states with transition rates designated by the $n\times n$ matrix, $G$, are governed by the rate equation, Eq.~\ref{eq:master}.
In principle, the system of equations can be solved analytically according to Eq.~\ref{eq:generalSoln}.
In practice, the dynamics can be efficiently simulated using a numerical ODE solver.

For a simulation of $g^{(2)}(\tau)$, the initial conditions are the state of the system immediately following the detection of a photon.
In a model with a single, radiative transition from excited $|e\rangle$ to ground state $|g\rangle$, and assuming background photons can be neglected (this is the case if the experimental $g^{(2)}(\tau)$ have been background-corrected), the initial condition is simply $P_{g}(0)=1$, with all other state populations equal to zero at time $t=0$.
The numerical solution of Eq.~\ref{eq:master} yields the time-dependent state populations, $P(t)$.
The steady-state populations, $P^{\infty}$ can also be found by numerically solving Eq.~\ref{eq:steadystate}.
Once the time-dependent and steady-state populations are found, quantities such as the PL intensity,
\begin{equation}
    I_\textrm{PL} = G_{ge}P_{e},
    \label{eq:PL}
\end{equation}
and $g^{(2)}(\tau)$ (Eq.~\ref{eq:autocorrelation})
can then be calculated.
The simulation can also be adapted to account for models involving multiple radiative transitions.
Details on the simulations in such cases can be found in Appendix~\ref{sec:apxSimu}.

This simulation tool can be linked with various physical models to compare simulations across changing experimental parameters such as excitation power and fields.
Power-dependent transitions, for example a pumped transition from a ground to excited state, can be incorporated by defining elements of a model's transition matrix to be dependent on a power parameter.
In a similar manner, electric or magnetic-field dependence of photon-statistics can be simulated by defining transitions that are a function of a field parameter.
For example, spin-dependent transition rates could change as a function of external magnetic field due to the system's spin Hamiltonian.

\begin{figure}
\includegraphics[trim={0 0 0 0},width = \linewidth]{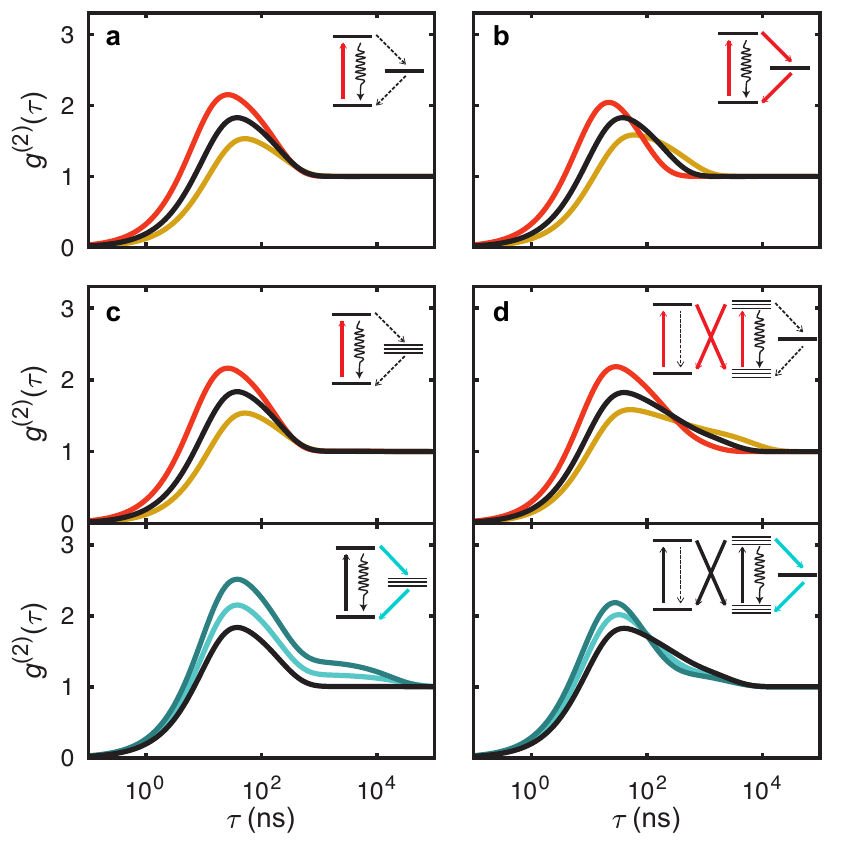}
\caption{\label{Figure5} Simulated $g^{(2)}$ traces for four different physical models. The effects of increasing (red) or decreasing (yellow) excitation power and a lower (light blue) or higher (dark blue) angle of magnetic field on $g^{(2)}(\tau)$ vary depending on the model. (a) Three-level model with a pumped transition (solid red arrow) to the excited state and fixed, non-radiative transition rates (dashed arrows) to a metastable state. (b) Three-level model with power-dependent transitions (solid red arrows) to and from a metastable state. (c) 5-level model with a pumped transition to the singlet excited state, and spin-dependent transitions (blue arrow) to and from a metastable spin triplet. (d) 9-level model of a nitrogen-vacancy center with both spin-dependent and power-dependent transitions. 
Radiative transitions are shown as wiggly arrows.
Subpanels in (c) and (d) show the effects of changing power (top subpanel) and magnetic field angle (bottom subpanel) separately for each model. Black curves are identical between top and bottom subpanels, and parameters are chosen such that the black curves for each model are qualitatively the same between (a-c) and approximate the amplitude and peak position of (d).}
\label{fig:simulations}
\end{figure}

Figure~\ref{fig:simulations} shows examples of simulated autocorrelation traces for four different models with varying excitation powers and magnetic fields.
The transition rates for each model were chosen so that the black curves in Figs.~\ref{fig:simulations}(a-c) are qualitatively the same as each other and similar to the black curve in the more complex model of Fig.~\ref{fig:simulations}(d).
However, the simulated $g^{(2)}$ curves for the models in Figs.~\ref{fig:simulations}(a-c) vary in qualitatively distinct ways as a function of excitation power and magnetic field.
Thus, in comparison with experimental PECS data, these simulations can be varied to help narrow down potential models.

Figure~\ref{fig:simulations}(a) depicts a basic three-level system, the simplest model that can host both antibunching and bunching dynamics.
The single radiative transition is denoted by the wiggly arrow.
A single power-dependent transition, designated by the solid red arrow, is varied in order to simulate $g^{(2)}(\tau)$ for high (red), medium (black) and low (yellow) excitation powers.

Figure~\ref{fig:simulations}(b) shows a three-level model similar to that in Fig.~\ref{fig:simulations}(a), but in this case the transitions to and from the metastable third state also depend on the excitation power.
This will be the case, for example, if the metastable state represents a different charge configuration than the ground and radiative excited state, which can be accessed through optically pumped ionization and recombination transitions.
Varying the excitation power differentiates between the models in Figs.~\ref{fig:simulations}(a) and \ref{fig:simulations}(b), as the bunching timescale $\tau_{2}$ changes dramatically as a function of power in \ref{fig:simulations}(b), in comparison to  Fig.~\ref{fig:simulations}(a) where the dominant power-dependent change is the bunching amplitude.
Experimental observations of $g^{(2)}(\tau)$ as a function of excitation power can be compared with such models to determine the nature of the non-radiative transitions and extract their scaling with optical excitation power.

In Fig.~\ref{fig:simulations}(c), spin dependent transitions are introduced to the model, represented by the blue arrow.
Here, varying a magnetic field angle that mixes the spin eigenstates of the metastable state can differentiate between the models in (a) and (c), whose traces exhibit similar power-dependent behavior.
The magnetic-field-dependent bunching dynamics (shown in blue) arise from the spin-dependent transitions in (c).
These spin-dependent transitions can sometimes be exploited to optically initialize and measure the spin state.
Hence, PECS measurements showing a variation in response to external magnetic fields can indicate the  presence of optically addressable spin states. 

As more is known about a system and its dynamics, PECS simulations can be extended to quite complex situations.
Figure~\ref{fig:simulations}(d), depicts a nine-level simulation of a nitrogen vacancy center, including both optically-driven ionization and recombination transitions as well as magnetic-field-dependent spin transitions.
A table of simulation parameters and additional details can be found in Appendix~\ref{sec:apxSimu}.

\section{PECS IN PRACTICE}

\begin{figure*}
\includegraphics[trim={0 0 0 0},width = \textwidth]{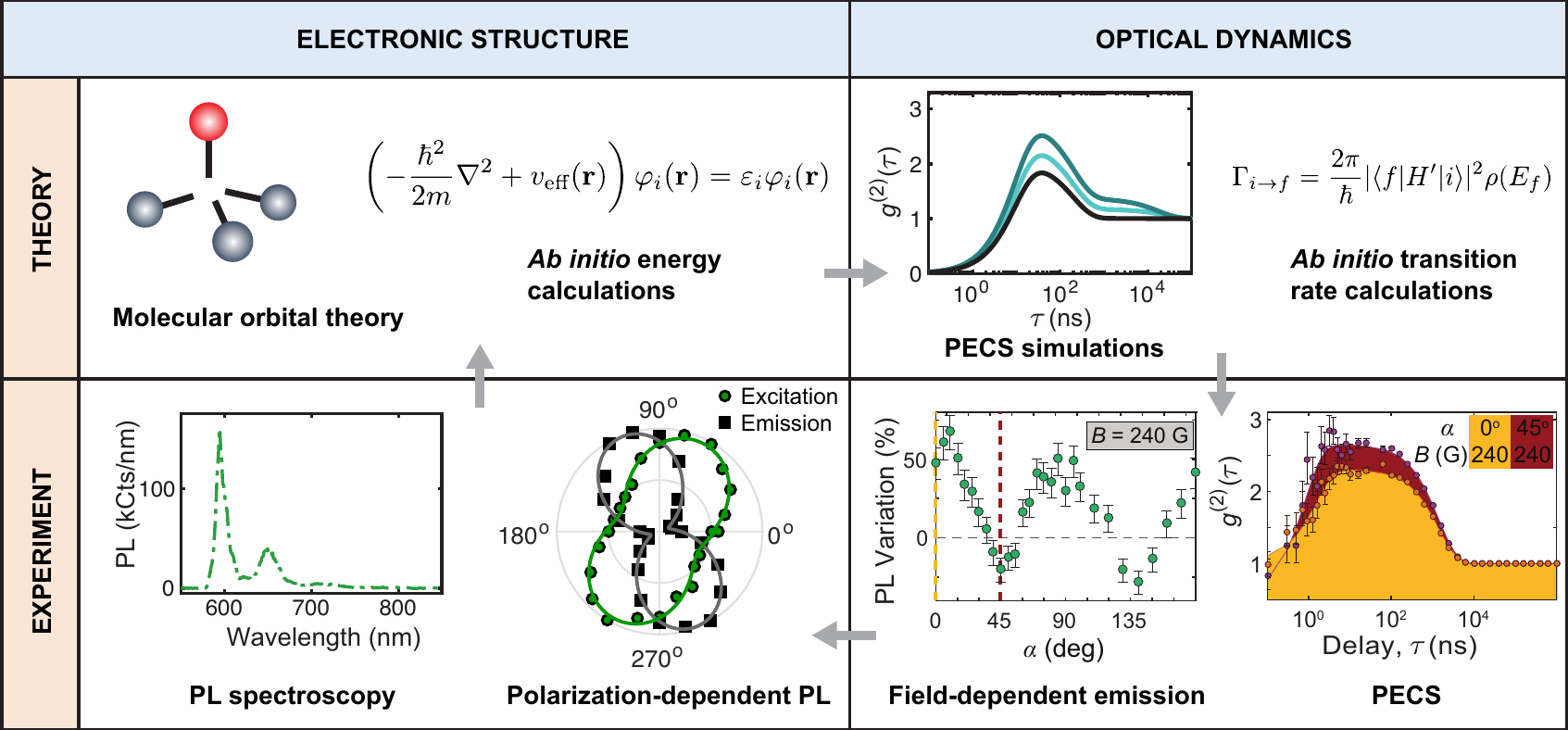}
\caption{Non-exhaustive table of tools to study quantum emitters. A variety of techniques, both experimental and theoretical, can give information about a quantum emitter's electronic structure and optical dynamics. Tools that help clarify electronic structure include experimental techniques, such as PL spectroscopy and polarization-dependent PL, and theoretical techniques of molecular orbital theory and \textit{ab initio} energy calculations. Tools that help uncover optical dynamics include theoretical techniques such as PECS simulations and \textit{ab initio} transition rate calculations, which should be supplemented with experimental techniques such as field-dependent emission and PECS measurements.  Images in the lower left quadrant depict hBN data adapted from Patel \textit{et al.} \cite{Pate}. Images in the lower right quadrant depict hBN data adapted from Exarhos \textit{et al.} 2017 \cite{Exar}.}
\label{fig:toolbox}
\end{figure*}

The methods outlined thus far for measurement, analysis, and simulation of $g^{(2)}(\tau)$ can be employed systematically alongside other tools to identify potential electronic models for unknown emitters.
Figure~\ref{fig:toolbox} illustrates several experimental and theoretical techniques that can be used in conjunction to deduce the electronic structure and optical dynamics of a quantum emitter.

Often the first step will be to characterize the basic optical properties of an emitter.
Typical measurements include PL emission spectra, autocorrelation to check for single-photon emission, and PL intensity as a function of excitation power to characterize saturation. These preliminary measurements can provide information about radiative transitions such as number of transitions, their lifetimes, and the strength of vibronic coupling, but they give little insight into non-radiative transitions.

Measuring the polarization-dependence of PL intensity can give further clarity on radiative transitions by revealing their symmetry. 
For polarization absorption measurements, the excitation polarization angle is varied, and the collected PL intensity is recorded as a function of excitation polarization angle.
For polarization emission measurements, the sample is excited at a fixed polarization, and a filter in the collection path is varied to filter the PL emission according to polarization angle.
These polarization measurements help clarify the number of dipole transitions and their orientations \cite{Neu2,Epst}.
Structured light beams with radial or azimuthal polarization can also reveal dipole orientation \cite{Taka}. 
Knowledge of dipole orientation with respect to the crystal axes can help single out potential point groups and can be a key piece of information when considering chemical models in cases where the defect chemistry is unknown \cite{Jung}.
Additional information about higher lying excited states can be gained through polarization measurements at various excitation wavelengths in relation to the zero-phonon line \cite{Roge,Jung2,Pate}.

The lower left quadrant of Fig.~\ref{fig:toolbox} depicts examples of these basic spectroscopic techniques applied to an emitter in hBN from Patel \textit{et al.} \cite{Pate}. In that work, the authors observed a single zero-phonon line and a phonon sideband consistent with a vibronic transition through a single optical dipole. This observation is further supported by the emission polarization measurement (black squares), which shows high polarization visibility. However, the measured PL intensity as a function of excitation polarization (green circles) is not aligned with the emission, and the visibility is reduced. 
Hence, the optical excitation and emission do not occur through the same optical dipole transition.
This measurement implies the presence of previously hidden excited states in the excitation pathway.

Initial characterization can also include measurements tailored toward specific properties of interest. For example, measurements that help identify optical spin signatures, such as magnetic-field-dependent PL can single out emitters with optically addressable spin states.
Any features of interest that are identified in initial characterization steps can be expanded upon through additional study including acquisition and analysis of photon emission statistics as a function of external fields.

Theoretical tools such as molecular orbital theory and \textit{ab initio} calculations can be applied together with experimental techniques to construct a baseline model for the emitter's electronic structure. The host material's crystal structure and its point groups constrain the types of level structures that can exist within the material. The symmetries of the crystal lattice, supplemented with information about the optical dipoles and from analyzed photon emission statistics, can help narrow down model parameters including the number of electronic levels, the number of spin or charge manifolds, and the characteristics of transitions. Density functional theory can point to likely defect chemistries through quantitative estimates of formation energies, which can be compared with experimental spectroscopic measurements of the emitter to predict the likelihood of different defect candidates. 

PECS measurements and simulations can be key to evaluating hypothesized models.  
Measuring $g^{(2)}(\tau)$ at different optical excitation powers and applied magnetic fields, and quantifying how the timescales and bunching amplitudes change accordingly, help unveil distinct dynamical processes. 
For example, Neu \textit{et al.} measured photon emission statistics at different excitation powers to help develop an electronic model for silicon vacancy centers in diamond \cite{Neu}.
Their observation of a bunching timescale with a nonlinear power dependence led them to suggest an excitation-power-dependent de-shelving process involving an additional excited state.
As another example, Patel \textit{et al.} proposed an indirect excitation mechanism to account for a nonlinear power-dependence of the antibunching rate observed for several hBN emitters \cite{Pate}. Using PECS simulations, Patel \textit{et al.} also clarified the effect of additional optically pumped transitions on the bunching rates and amplitudes. These simulations allowed the authors to distinguish between emitters with metastable states accessed through optically pumped or spontaneous transitions.
In general, power-dependent PECS measurements can reveal the presence of non-radiative states and their associated lifetimes.

Meanwhile, PECS measurements as a function of externally applied magnetic fields can reveal the energetics and dynamics of spin states.
As an illustrative example, the lower right panel of Fig.~\ref{fig:toolbox} shows magnetic-field-dependent PL and PECS data from an hBN emitter observed by Exarhos \textit{et al.} \cite{Exar}.
In this case, the steady-state PL variations in response to applied dc magnetic fields suggested the presence of spin states and spin-dependent optical transitions.
Field-dependent PECS measurements revealed that the decreases in PL were correlated with increases in bunching amplitude, but with no change in bunching timescale.
Molecular orbital theory also provided some insight into the emitter's electronic structure. 
Using selection rules from the defect's symmetry group, the authors narrowed down the possible models to two options, which they were able to distinguish between through PECS simulation of both models.
As a result, Exarhos \textit{et al.} showed that the magnetic-field dependence of autocorrelation bunching amplitudes and timescales was consistent with a spin-dependent intersystem crossing.
Simulated PECS data can be empirically fit using Eq.~\ref{eq:empirical_g2} to quantitatively compare timescales and bunching amplitudes to those observed in experimental data.
In addition, the measured PL data can be compared with simulations as a function of optical power or applied field.
Dynamical information gained through PECS measurements and simulation can be further supplemented through \textit{ab initio} calculations, which can give quantitative estimates of vibronic coupling strengths, electron capture rates, ionization cross sections, and nonradiative transition rates \cite{Alka, Gali}.


The process of studying a quantum emitter using the tools in Fig.~\ref{fig:toolbox} should be iterated until there is enough experimental information to support a particular proposed model, and simulations can reproduce similar phenomena to those observed.
The result may still be an approximation of the true underlying model. 
However, in revealing key properties of the emitter, the outcome of the combined experimental and theoretical approaches can provide enough of a foundation to begin to realize applications.

Formulating a structural and dynamical model of a quantum emitter is crucial to harnessing properties for quantum technologies.
Deeper understanding of a system allows its strengths to be connected with particular applications. 
For example, emitters hosting excited states with short optical lifetimes may be useful as single photon sources or in applications that require high signal-to-noise ratio such as quantum communication.
Conversely, long-lived electronic spin states can serve as quantum memories for applications that require the storage of quantum states such as quantum registers for quantum repeaters.
Magnetic-field-sensitive transitions can be utilized for quantum sensing.
With a more complete idea of the emitter's properties and optical dynamics, it becomes possible to evaluate how its strengths and weaknesses compare to those of existing platforms. 

\section{CONCLUSION}
PECS is an easy-to-implement experimental technique, but it remains under-utilized. 
In this tutorial, we have discussed the theoretical foundation and application of PECS for quantum emitters, highlighting the potential of PECS to reveal optical dynamics when supplemented with other spectroscopic techniques. 
While standard optical characterization techniques provide information about radiative transitions, the non-radiative transitions, which PECS is particularly suited to resolve, are often those whose properties are leveraged for various quantum technologies.

With proper attention to acquisition, analysis, and interpretation, PECS can provide detailed information about a quantum emitter's electronic structure and dynamics that allows for the design of efficient quantum control protocols. Here, we have discussed a number of tools for the community to use for acquisition, analysis, simulation, and interpretation of PECS.
Expanding the set of available quantum emitters and host materials, each with specific advantages, will lead to continuous advances in science and technology.

\section*{ACKNOWLEDGEMENTS}

The authors thank S. Thompson, J. Gusdorff, and M. Ouellet for helpful discussions. 
This work was supported by the National Science Foundation under award DMR-1922278.

\begin{appendices}
\appendix
\def\appendixname{APPENDIX}
%
%
%
%
%
%
%



\section{WAITING TIME ADDITIONAL DERIVATIONS}\label{sec:apxWait}

This section contains additional derivations of $W(\tau)$. The first deals with the effect of imperfect collection efficiency on $W(\tau)$, and the second discusses an example of deriving $W(\tau)$ for models with $n>2$ levels. In particular, we explore the cases of a two-level system with imperfect collection efficiency and a three-level system with one non-radiative transition and perfect collection efficiency.

\subsection{Collection-efficiency dependence}



The deriviation of the waiting time distribution in Eqs.~\ref{eq:waittime}-\ref{eq:waittime2lvl} assumes the condition of perfect collection efficiency.
However, in a realistic experiment, the setup collection efficiency, $C$, significantly impacts the probability of receiving consecutive photons.
Thus, a derivation of $W(\tau)$ that accurately captures experimental realities must incorporate $C$. For simplicity, we show the derivation for a two-level system.

As with perfect collection efficiency, we consider the probability that the system starts in the ground state, evolves to the excited state, then decays back to the ground state after delay $\tau$, emitting a photon. 
However, in this case the system can evolve through any number of cycles between excited and ground before the detection of a subsequent photon. Additional multiplicative factors, $C$ and $1-C$, account for the respective probabilities that the subsequent photon is detected or is not detected once it is emitted, and we integrate and sum over all possible combinations of losing $n$ photons before detecting the next photon.
This is equivalent to the infinite sum of convolutions,
\begin{equation}
\begin{split}
    W(t) &= \frac{C}{1-C}\biggr((1-C)P_{g\rightarrow e}(t)*P_{e\rightarrow g}(t) + \\
    &(1-C)^{2}P_{g\rightarrow e}(t)*P_{e\rightarrow g}(t)*P_{g\rightarrow e}(t)*P_{e\rightarrow g}(t)\\
    &+\cdots\biggl),
\end{split}
\label{eq:waitC}
\end{equation}
where $P_{a\rightarrow b}(t)$ is given by Eq.~\ref{eq:pdef}.

Equation~\ref{eq:waitC} can be evaluated in Laplace space following the general relation for an infinite sum of convolutions of the same function, $h(t)$,
\begin{equation}
\begin{split}
    h(t)+h(t)*h(t)+\cdots = \mathcal{L}^{-1}\left\{
    \frac{\mathcal{L}\{h\}(s)}{1-\mathcal{L}\{h\}(s)}\right\}(t),
    \label{eq:laplace}
\end{split}
\end{equation}
where $\mathcal{L}$ is the Laplace transform and $s$ is a complex frequency parameter.

Therefore, defining $h(t)$ as
\begin{equation}
    h(t) = (1-C)P_{g\rightarrow e}(t)*P_{e\rightarrow g}(t),
\end{equation}
the probability that the system evolves but a photon is not received, yields the collection-efficiency dependent expression for $W(\tau)$ for a two-level system,
\begin{equation}
\begin{split}
    W(\tau) &= \frac{2C\Gamma_{eg}\Gamma_{ge}}{\sqrt{-4C\Gamma_{eg}\Gamma_{ge} + (\Gamma_{eg} + \Gamma_{ge})^{2}}}e^{-\frac{\Gamma_{eg}+\Gamma_{ge}}{2}\tau}\\
    &\sinh{\frac{1}{2}\sqrt{-4C\Gamma_{eg}\Gamma_{ge} + (\Gamma_{eg} + \Gamma_{ge})^{2}}\tau}.
    \end{split}
    \label{waittimeCollEff}
\end{equation}
This relation between $C$ and $W(\tau)$ for a two-level model is illustrated in Fig.~\ref{fig:waitingTime}.


\subsection{Systems with $n>2$ levels}
Compared to $g^{(2)}(\tau)$, which is more straightforward to generalize from a two-level to a multi-level model, $W(\tau)$ requires a unique derivation for each specific electronic model and can be difficult to evaluate.
Here we consider the most basic example of a three-level model with both a radiative and non-radiative pathway to the ground state and perfect collection efficiency.
As with the derivation for imperfect collection efficiency, there is a need to account for all cases where the system evolves to the ground state, but a photon is not detected.
In this case, a delay in receiving a subsequent photon stems from the non-radiative transition through a third, metastable state to the ground state.
The system can evolve through any number of non-radiative cycles prior to the emission of a subsequent photon.


Starting with Eq.~\ref{eq:WaitGeneral}, we define the probability density function of a full non-radiative cycle from ground state to excited to metastable back to ground as,
\begin{equation}
    h(t) = P_{g\rightarrow e}(t)*P_{e\rightarrow m}(t)*P_{m\rightarrow g}(t),
\end{equation}
where the subscript, $m$, indicates the metastable third state.
The waiting time distribution for a three-level model is then given by,
\begin{equation}
    W(t) = P_{g\rightarrow e}(t)*P_{e\rightarrow g}(t)(1+ h(t)+h(t)*h(t)+\cdots),
    \label{eq:WaitTime3lvl}
\end{equation}
which can be evaluated using Eq.~\ref{eq:laplace}.
Additional model complexity such as other nonradiative decay pathways or imperfect collection efficiency further complicate the derivation of $W(\tau)$. However, such features are common in realistic models. As a result, autocorrelation presents a more tractable tool for measuring the optical dynamics of realistic models.

\section{AUTOCORRELATION FROM MULTIPLE RADIATIVE TRANSITIONS}\label{sec:apxMultRad}

In the main text, we simplify the discussion by limiting it to models with a single radiative transition. While this suffices in many situations, it also is common for systems to have multiple radiative transitions whether they be due to different spin states or charge manifolds or some other mechanism. In all cases the transition rate matrix, $G$, still determines the state evolution according to the rate equation, Eq. ~\ref{eq:master}. However, the initial condition, $P_{0}$, which is the state immediately following photon emission, is dependent on which transitions are radiative. Similarly, optical dynamics such as intensity are also affected, which in turn impacts the autocorrelation function.
In order to account for the effect of multiple radiative transitions on $g^{(2)}(\tau)$, we introduce the transition collection efficiency matrix, $C$. $C$ is made up of individual elements, $C_{ij}$, that give the probability of detecting a photon from each transition from state $j$ to state $i$. $C_{ij}$ can take on values from 0 to 1 with fractional values accounting for different collection efficiencies for different transitions, which might arise from polarization selection rules or different emission wavelengths.



With the collection efficiency matrix defined, PL can be calculated. The steady-state rate at which photons are detected from a transition $j$ to $i$ ($I_{ij}$) depends on the steady-state population of state $j$ ($P_{j}^{\infty}$), the transition rate from the state $j$ to state $i$ ($G_{ij}$), and the probability of collecting a photon from that transition ($C_{ij})$:
\begin{equation}
    I_{ij} = C_{ij}G_{ij}P^{\infty}_{j}.
\end{equation}

The total steady-state photoluminescence ($I_{PL}$), which is the average rate at which photons are detected from any transition, is given by summing over the photon detection rates from all transitions,
\begin{equation}
    I_{PL} = \sum_{ij}^{n}I_{ij}.
\end{equation}

Therefore the probability of detecting a photon from a specific transition from $j$ to $i$ ($P_{\gamma}^{\infty (j\rightarrow i)}$) is given by the fractional contribution of that transition to the total PL:
\begin{equation}
    P_{\gamma}^{\infty (j\rightarrow i)} = \frac{I_{ij}}{I_{PL}} = \frac{C_{ij}G_{ij}P^{\infty}_{j}}{\Sigma_{ij}^{n}C_{ij}G_{ij}P^{\infty}_{j}}.
\end{equation}

As a result, the distribution of states following the detection of the photon is a column vector, $P_{0}$, with components given by the probability of detecting a photon from any transition into state $i$,
\begin{equation}
    P_{0}^{i} = \sum_{j}^{n} P_{\gamma}^{\infty (j\rightarrow i)} =  \sum_{j}^{n}\frac{C_{ij}G_{ij}P^{\infty}_{j}}{\Sigma_{ij}^{n}C_{ij}G_{ij}P^{\infty}_{j}}.
    \label{eq:P0}
\end{equation}

As discussed in the main text, the autocorrelation function is proportional to the probability of receiving any photon at time $t_{2}$, given one was received at time $t_{1}$.
Therefore, writing $g^{(2)}(\tau)$ for multiple radiative transitions requires accounting for the time-dependent populations of all radiative states and the transition rates and collection efficiency of the transitions out of those states. With the initial state given by Eq.~\ref{eq:P0}, and properly normalized to the steady-state, this gives the equation for autocorrelation from multiple radiative transitions:
\begin{equation}
    g^{(2)}(\tau = t_{2}-t_{1})=\frac{\Sigma_{i}^{n}\Sigma_{j\neq i}^{n}C_{ij}G_{ij}P_{j}(t_{2}|P(t_{1}) = P_{0})}{\Sigma_{i}^{n}\Sigma_{j\neq i}^{n}C_{ij}G_{ij}P_{j}^\infty}.
\end{equation}

In the case of a single radiative transition, this reduces to Eq.~\ref{eq:autocorrelation}
Autocorrelation data from emitters with multiple radiative transitions can still be captured by the empirical fit, Eq.~\ref{eq:empiricalFit}, with the possibility that multiple radiative transitions could lead to multiple antibunching timescales.

\section{ALGORITHM FOR PROCESSING PHOTON CORRELATIONS}\label{sec:apxAlg}

\begin{figure}
\includegraphics[trim={0 0 0 0},width = \linewidth]{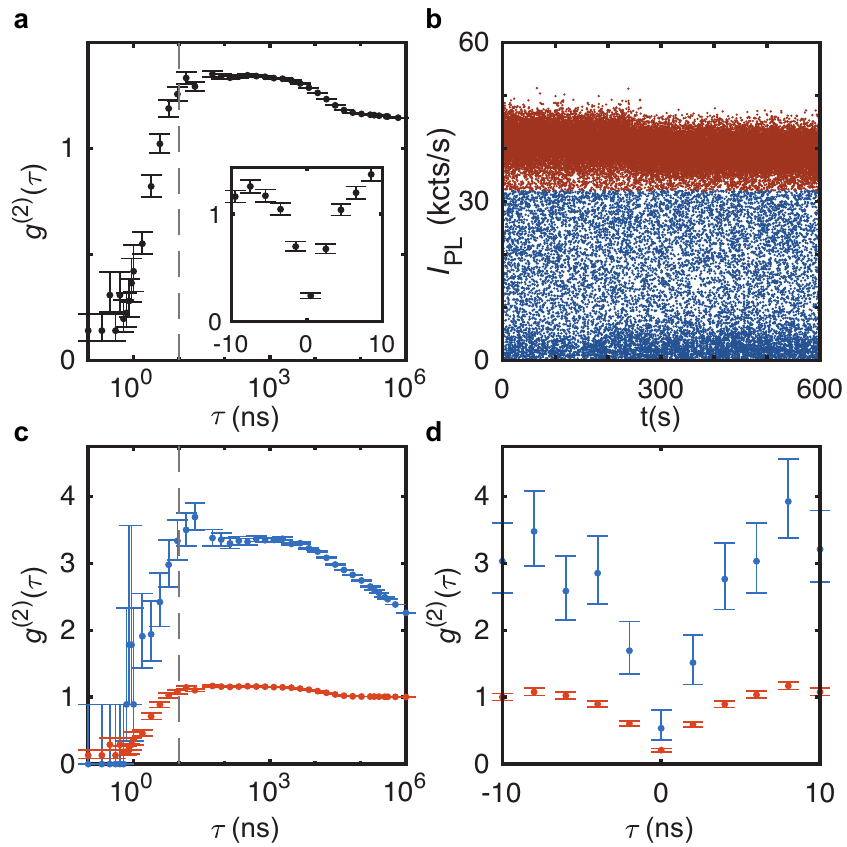}
\caption{Output of time-tagged-time-resolved data processed for photon correlations shown for an example case of an emitter in hBN with blinking.
(a) Raw photon emission statistics data from an emitter in hBN processed with logarithmic bins. Inset shows the same data processed on a linear scale. Dashed line denotes $\tau$=10ns.
(b) Intensity vs time data binned at 0.01s displaying partitioning at 32kcts/s between an on (orange) blinking state and off (blue) blinking state.
(c) Photon emission statistics data from the same photon time series as (a) processed on a logarithmic scale based on intensity thresholding in (b). Bunching dynamics are mainly present in the photon emission statistics of the dark state (blue), whereas the data processed for the bright state (orange) adheres more closely to a two-level model.
(d) Photon emission statistics data processed on a linear scale based on the intensity thresholding in (b). }
\label{fig:g2codeEx}
\end{figure}

The time-tagged data acquired by two detectors and a time correlated single-photon counter (TCSPC) consists of two separate time series detailing the arrival times of all photons detected by each detector during acquisition.
Photon statistics processing code and example can be found at https://github.com/penn-qel/photon-emission-correlation-spectroscopy.
The processing algorithm \texttt{TTTR\_cross\_correlation(parsedData,options)} based on \emph{Laurence et al.} \cite{Laur}, uses the input of two time series along with various processing parameters to calculate the normalized correlations and their uncertainties. The algorithm uses binary search functions defined within it to bin and count correlations thus avoiding the need to iterate through all individual events and significantly reducing processing time. Additional features include processing of average count rates to track stability during measurement and an option for partitions to process correlations between different blinking states separately.

\subsection{Preparing the data}
Certain TCSPCs store photon time series in particular file formats, which must be parsed before running the processing algorithm. The function \texttt{TTTR\_import\_PTU(filenames)} takes time series stored in a .ptu file format and stores the data and metadata as a MATLAB structure, \texttt{TTTRData}, which is then fed to the function \texttt{TTTR\_extract\_channel\_times(TTTRData,options)} to convert the .ptu data into a MATLAB array of times. These are output into the \texttt{parsedData} structure containing the times and total number of counts for each channel, global resolution, and total acquisition time.

\subsection{Defining processing parameters}
The \texttt{options} structure allows the user to define processing parameters. These set the range (\texttt{tauLimits}) and resolution (\texttt{tauRes}) over which the correlations will be calculated. Resolution should be chosen such that it is smaller than the fastest timescale desired for measurement. Since decreasing resolution also increases uncertainties, which are Poissonian and depend on the number of counts per bin, scans that require lower resolution may require greater acquisition time. The choice of resolution and limits also affect the processing time, so a balance must be struck between resolution, magnitude of uncertainties, and processing time. 

Other parameters in the options structure include \texttt{tRes} to specify the resolution for count rates calculations,\texttt{tauAxis} to choose specific axis over which cross-correlation will be calculated, \texttt{countRateRanges} to partition blinking data based on intensity, \texttt{tFlag} to partition the time axis for counts rate calculation, and \texttt{verbose} and \texttt{statusbar} to enable command window and pop-up updates.

\subsection{Algorithm Detail}
The algorithm first constructs two time axes, \texttt{tauAxis} and \texttt{tAxis}, defining the bins over which the correlations and average counts rates will be calculated respectively.  The delay time axis, \texttt{tauAxis}, is constructed from the processing parameters defined in the options structure. \texttt{tAxis} is constructed based on the total acquisition time and resolution, \texttt{tRes}. 
Count rates in each channel are calculated by counting the number of events in each bin that \texttt{tAxis} defines and dividing by bin width.
The overall count rate for each channel, \texttt{avgRate}, is calculated from the channel's total counts and total measurement time in order to compute the normalization factor for $g^{(2)}(\tau)$. (See Sec.~\ref{sec:DataProc}).

To process the cross correlations between the two channels, the algorithm iterates through each time-tagged photon event in the channel with fewer events, referred to as Channel 1. With each iteration, the zero-delay reference point for the bins defined by \texttt{tauAxis} is updated according to the time of the current photon event, $t_{0}$. Then, events in Channel 2 are binned according to \texttt{tauAxis}, with $\tau = 0$ corresponding to $t_{0}$, and the number of photons that fall within each bin in Channel 2 are counted as the correlations.




\subsection{Outputs}
The processing algorithm outputs a structure, \texttt{T2data}, that contains the axis of delay times, raw correlated data, normalized $g^{(2)}(\tau)$, and errors. If count rates are calculated, the output structure will also include the time axis for count rates, the average count rates over acquisition time, and flags for partitioning the time axis. Figure~\ref{fig:g2codeEx} contains examples of processing code outputs along with a demonstration of several features of the code including log and linear binning options, calculation of count rates, and separate processing of different blinking states.

\subsection{Additional complexity}
The algorithm offers options of both log and linear processing of the correlations with examples of the resulting outputs shown in Fig.~\ref{fig:g2codeEx}(a) and inset respectively. Linear binning allows for resolution of features at low times and log binning allows for resolution of timescales at both short and long times. Log binning is achieved by pre-defining bins with logarithmic spacing prior to calculating correlations.

The processing algorithm also contains functionality for processing different blinking states of an emitter to extract state-specific optical dynamics. Time partitions can either be pre-defined with flags using \texttt{tFlag} in the options structure or can be calculated based on intensity thresholds specified by \texttt{countRateRanges} in the options structure. In the latter case, the specified threshold for the partitions is applied to the time-dependent intensity data as demonstrated in Fig.~\ref{fig:g2codeEx}(b) where a threshold of 32kcts/s separates a bright state at $\sim$40kcts/s from a dark state at $\sim$3kcts/s. The correlations for each state can then be calculated separately resulting in two different autocorrelation traces shown in Fig.~\ref{fig:g2codeEx}(c) on a logarithmic scale and Fig.~\ref{fig:g2codeEx}(d) on a linear scale.


In the processing algorithm, upper and lower errors ($\Delta^{\pm}$) are calculated for each bin according to Poisson error with asymmetric errors,
\begin{equation}
    \Delta^{\pm} = \frac{\sqrt{M+\frac{1}{4}}\pm\frac{1}{2}}{I_{\textrm{A}}I_{\textrm{B}}wT},
\end{equation}
for bins with close to zero counts at which point Poisson error (Eq.~\ref{eq:poissonErr}) gives inaccurate results \cite{Barl},

\section{AKAIKE INFORMATION CRITERION}\label{sec:apxAkai}

When analyzing experimental data for newer or unknown emitters, the electronic level structure and dynamics of the emitter are often unknown. As a result, a fitting routine often must be implemented to determine which empirical model best fits the data and its features.
The Aikaike Information Criterion (AIC) can be used as a measure of relative quality to identify a model that captures the dynamics comparatively best. In order to implement it, the data is fit to several models of Eq.~\ref{eq:empiricalFit}, varying $n$, the number of electronic levels. The AIC is given by

\begin{equation}
    \textrm{AIC} = 2p - 2\textrm{ln}(L).
\end{equation}

For a nonlinear fit with normally distributed errors,
\begin{equation}
    \textrm{ln}(L) = 0.5\left(-N\left(\textrm{ln}(2\pi)+1-\textrm{ln}(N)\textrm{ln}\left(\sum_{i=1}^{n}x_{i}^{2}\right)\right)\right),
\end{equation}

where $N$ is the total number of data points and $x_{i}^{2}$ are the residuals.

In order to compare models, the \textrm{AIC} for each model is calculated and compared. The model with the lowest \textrm{AIC} representing the model most likely to be correct, while the relative likelihood for the other models can be calculated as
\begin{equation}
    \textrm{exp}\left(\frac{\textrm{AIC}_\textrm{min}-\textrm{AIC}_{i}}{2}\right).
\end{equation}

AIC can be supplemented with other measures of fit such as reduced chi-squared to empirically determine the best choice of $n$.


\begin{table*}[t]
    \renewcommand{\thetable}{S\arabic{table}}
    \renewcommand\tablename{Table}
    \caption{\textbf{Simulation parameters for Fig~\ref{fig:simulations}}}    
    \label{tab:SimParams}
    \begin{tabularx}{1\linewidth}
    { | X | X | X | X | X | }
    \rowcolor{lightGray}[\tabcolsep]
    \hline
     \textbf{Parameter} & \textbf{3-level model (spontaneous transitions)} & \textbf{3-level model (pumped transitions)} & \textbf{5-level model} & \textbf{Nitrogen-Vacancy model}\\
    \hline
    $k_{\textrm{ex}}$ (MHz)  & [25,50,100] & [25,50,100] & [25,50,100] & [13.125,26.25,52.5] \\
    \hline
    $k_{\textrm{r}}$ (MHz) & 50 & 50 & 50 & 75 \\
    \hline
    $k_{\textrm{isc0}}$ (MHz) & 5 & $5k_\textrm{ex}$  & 4.9995 & 5 \\
    \hline
    $k_\textrm{isc0,out}$ (MHz) & 2.5 & $2.5k_\textrm{ex}$  & 2.5 & 3.11 \\
    \hline
    $k_{\textrm{isc}\pm}$ (MHz) & - & - & 2.5E-4 & 60 \\
    \hline
    $k_{\textrm{isc,out}\pm}$ (MHz)  & - & - & .025 & 2.75 \\
    \hline
    $k_{\textrm{ion}}/k_{\textrm{rec}}$  & - & - & - & $3k_\textrm{ex} / 2.25k_\textrm{ex} $ \\
    \hline
    $\mathbf{B}$ amplitude (G) & - & - & 46 & 300 \\
    \hline    
    $\mathbf{B}$ angle (deg)  & - & - & [0,30,60] & [0,15,50] \\
    \hline    
    \end{tabularx}
\end{table*}

\section{OPTICAL DYNAMICS SIMULATION}\label{sec:apxSimu}
The optical dynamics simulation \texttt{simulate\_autocorrelation(ModelPars,SimPars,Opts)}, takes the input of an electronic model defined by states, the transition rates between them, and the photon collection efficiencies of the transitions and returns the populations of each state given an initial condition. Optical dynamics such as the photon statistics and photoluminescence are also calculated. The simulation is executed through MATLAB’s ODE solver  \texttt{ode15s}.
 Simulation code and example can be found at https://github.com/penn-qel/photon-emission-correlation-spectroscopy.
\subsection{Defining the model}
The electronic model is defined through three inputs: The number of levels (\texttt{nLevels}), the transition rates between levels ($G$), and the collection efficiency matrix ($C$) all of which are passed to the function through the \texttt{ModelPars} or \texttt{SimPars} structure. The generator matrix, $G$, defines the transition rates between levels and takes the form of an nLevels x nLevels matrix with off-diagonal elements $G_{ij}$ giving the transition rates from state $|j\rangle$ to state $|i\rangle$, and diagonal elements $G_{ii}$ giving the negative sum of all rates leaving state $|i\rangle$. The columns of $G$ all sum to 0. The collection efficiency matrix, $C$, is an nLevels x nLevels matrix with elements $0\leq C_{ij}\leq1$ denoting the fractional probability of collecting photons from the transition from state $|j\rangle$ to state $|i\rangle$. A non-radiative transition would be denoted by $C_{ij} = 0$ while a radiative transition would have $C_{ij} = \epsilon$ where $\epsilon$ is the collection efficiency from that transition.

\subsection{Simulation Detail}
The steady-state populations are calculated from $G$ using MATLAB’s \texttt{null()} function and normalized such that the sum of all steady-state populations is 1.
 PL is calculated from the steady-state populations and $C$ matrix (see Eq. ~\ref{eq:PL}).
The eigenrates of $G$ are calculated through the MATLAB function \texttt{eig()} and ordered and identified as real or imaginary rates. The min and max eigenrates are used to determine a range of the time values ($t$) to input to the ODE solver. The initial conditions are set according to Eq.~\ref{eq:P0}.  The ODE solver \texttt{ode15s} is run using the inputs of the rate equation, generator matrix, and initial conditions.  
A comparison between the steady-state populations calculated through the null vector and through the ODE solver can be used as an estimate of simulation error, $p_\textrm{err}$.

\subsection{Simulation Outputs}
The steady-state populations, steady-state PL, eigenvalues of $G$, time-dependent populations of states, and vectors of $g^{(2)}$ and $t$ are all returned in the \texttt{SimPars} structure.

\subsection{Additional complexity}
Multiple simulations can be executed by feeding multiple pairs of $G$ and $C$ matrices defining different models or model conditions to the \texttt{SimPars} structure. The number of elements in \texttt{SimPars} determines the number of simulations that will be run. Transition rates dependent on physical interactions with fields such as spin or charge phenomena can be defined prior to execution of the simulation such that the model incorporates additional phenomena.
The transition rates between states can be defined as spin states that are dependent on an applied magnetic field, $B$. This takes the form of defining the spin-field interaction through the Hamiltonian, and calculating how an applied field leads to a change of basis, represented by a change transition rates from spin-dependent states.

\subsection{Simulation detail for Fig.~\ref{fig:simulations}}
Curves in Fig.~\ref{fig:simulations} were simulated using the method described above. Simulation parameters used to generate Fig.~\ref{fig:simulations} can be found in Table \ref{tab:SimParams}.
Here, the excitation rate is given by $k_{\textrm{ex}}$, the emission rate is given by $k_\textrm{r}$, and the rates to and from the the inter-system crossing (isc) metastable state are $k_\textrm{isc0}$ and $k_\textrm{isc0,out}$ respectively. For models with spin states in the metastable state, such as the 5-level model and NV model, $k_\textrm{isc0}$ ($k_\textrm{isc0,out}$) specifies the zero-$B$-field transition rate to (from) the $m_S=0$ spin-triplet sublevel, while $k_{\textrm{isc}\pm}$ ($k_{\textrm{isc,out}\pm}$) specifies the zero-$B$-field transition rate to (from) the $m_S=\pm1$ spin sublevels. $k_\mathrm{ion}$ and $k_\mathrm{rec}$ are the ionization and recombination rates to and from NV$^{0}$ and NV$^{-}$. The $B$-field angle is given with respect to the defect axis. Power-dependent transitions are shown as a multiple of $k_{\textrm{ex}}$. For models with spin-dependence (5-level and NV), the Hamiltonian specified is for the case of a spin-1 triplet configuration with a single symmetry axis and takes the form,
\begin{equation}
    H = g\mu_{B}\mathbf{B}\cdot\mathbf{S} + D(S_{z}^{2}-\frac{1}{3}S(S+1))
\end{equation}
where $g$ is the isotropic $g$-factor, $\mu_{B}$ is the Bohr magneton, $\mathbf{B}$ is the magnetic-field vector, $D$ is the zero-field splitting, and $\mathbf{S}$ and $S_{z}$ are spin-1 operators.
For the 5-level model simulation, we assume $D=1000$~MHz for the spin-triplet metastable state. For the NV center, the values used were $g_\textrm{ES} = 2.01$, $g_\textrm{GS} = 2.0028$, $D_{ES} = 1425$ MHz, and $D_{GS} = 2859$ MHz, where ``ES'' and ``GS'' refer to the excited and ground state respectively.
When a magnetic field is applied to a model with spin states, the interaction between the spin states and the field, as defined by the Hamiltonian, results in spin mixing.
In the NV model, the inner product of the excited state and ground state eigenvectors adjusts the baseline excitation ($k_{ex}$) and emission ($k_{r}$) rates to give the distribution of excitation and emission rates between the ground and excited spin states.
Similarly, for both the NV and 5-level model, other spin-dependent transition rates such as the inter-system crossing rates ($k_\textrm{isc}$) specified in Table~\ref{tab:SimParams} are adjusted according to the new calculated spin projections in a magnetic field.





%

\end{appendices}


\bibliography{PESPaperRefs4}

\end{document}